\documentclass[12pt, letterpaper]{JHEP3}

\usepackage{epsfig}

\usepackage{amsmath}

\usepackage{graphics}

\def\makeatletter{\catcode`\@=11}
\makeatletter
\def\mathbox#1{\hbox{$\m@th#1$}}%
\def\math@ccstyles#1#2#3#4#5#6#7{{\leavevmode
      \setbox0\mathbox{#6#7}%
      \setbox2\mathbox{#4#5}%
      \dimen@ #3%
      \baselineskip\z@\lineskiplimit#1\lineskip\z@
      \vbox{\ialign{##\crcr
             \hfil \kern #2\box2 \hfil\crcr
             \noalign{\kern\dimen@}%
             \hfil\box0\hfil\crcr}}}}
\def\mathaccstyles{\math@ccstyles\maxdimen}
\def\maththroughstyles{\math@ccstyles{-\maxdimen}}
\def\unity%
 {\maththroughstyles{.45\ht0}\z@\displaystyle {\mathchar"006C}\displaystyle 1}

\title{Shocks and Universal Statistics in (1+1)-Dimensional Relativistic Turbulence}

\author{Xiao Liu and Yaron Oz
\\

~\\
Raymond and Beverly Sackler School of Physics and Astronomy \\
Tel-Aviv University, Ramat-Aviv 69978, Israel\\

\vspace{0.8cm}

\email{xiao.liu.tau@gmail.com, yaronoz@post.tau.ac.il}

}

\abstract{We propose that statistical averages in relativistic turbulence exhibit universal properties. We consider analytically the velocity and temperature differences structure functions in the (1+1)-dimensional relativistic turbulence in which shock waves provide the main contribution to the structure functions in the inertial range. We study shock scattering, demonstrate the stability of the shock waves, and calculate the anomalous exponents.
We comment on the possibility of finite time blowup singularities.
}

\preprint{TAU-10-xxx}

\newcommand{\pd}{\partial}

\def\be{\begin{equation}}

\def\ee{\end{equation}}

\def\bea{\begin{eqnarray}}

\def\eea{\end{eqnarray}}

\begin{document}

\tableofcontents

\section{Introduction and Summary}

While turbulent flows are rather generic in nature,
an analytical understanding of the flows in the non-linear regime is still lacking.
The evolution of fluid flows at small Mach number, where the characteristic velocity is much smaller than the speed of sound, is
described by the incompressible Navier-Stokes equations for the velocity field $v_i(x,t)$ and the pressure $p(x,t)$.
The velocity field exhibits a  highly complex spatial and temporal structure in the non-linear regime, which is very senstitive
to small changes in the initial conditions and the external forcing.
Thus, instead of considering a single realization of the velocity field, one considers the statistics of velocity differences between points separated by a fixed distance, defined by averaging over space or with respect to an external random force.

At large Reynolds number, there is a postulated range wherein the nonlinear advection cascades energy through scales with no
dissipation until viscosity sets in at the dissipation scale.
In this so-called inertial range, the statistical properties of the flow are postulated to become homogeneous and isotropic.
There is numerical and experimental evidence that the statistical averages
exhibit universal properties in the inertial range, which one hopes to study analytically.
For instance, the longitudinal structure functions of order $n$, (${\bf r} \equiv {\bf x} - {\bf y}$)
\begin{equation}
S_n(r) \equiv \left\langle \left(({\bf v}({\bf x})-{\bf v}({\bf y}))\cdot \frac{{\bf r}}{r}\right)^n\right\rangle \ ,
\end{equation}
exhibit a universal behavior where
\begin{equation}
S_n(r) \sim r^{\xi_n} \ ,
\label{expon}
\end{equation}
and $\xi_n$ is a universal function independent of the statistics of the forcing.

When $n=3$ and with three spatial dimensions one has the
Kolmogorov's four-fifths law for incompressible Navier-Stokes turbulence
\begin{equation}
S_3(r)=-\frac{4}{5}\epsilon r \ . \label{Kolm}
\end{equation}
Here $\epsilon$ is the mean rate of energy dissipation per unit volume due to viscosity.
This law holds, within the inertial range, regardless of detailed properties of the fluid, the initial and boundary physical conditions on the fluid, and the energy pumping mechanisms on the large scales \cite{Kolmogorov}.
Until recently, the Kolmogorov four-fifths law has been the only non-trivial known result on the statistics of incompressible turbulence.

It has been observed in \cite{FIO,FO2} that the Kolmogorov's four-fifths law does not require the cascade picture for its derivation,
but simply follows from the fact that the Navier-Stokes equation is a conservation law.
This allowed a derivation of  new exact scaling relations in nonrelativistic
and relativistic turbulence \cite{FIO,FO2}.
The Kolmogorov law
turns out to be the low Mach number limit of a general statistical law holding for compressible turbulence with finite Mach number \cite{FIO}.
In relativistic  hydrodynamics one derives exact scaling relations of energy-momentum tensor two-point functions \cite{FO2}
\begin{equation}
\langle T_{0j}(x, t) T_{ij}(y, t)\rangle=\frac{\epsilon}{d} |x_i-y_i| \ ,
\label{stress}
\end{equation}
where $d$ is the number of space dimensions, $\epsilon$ is a constant that depends on the forcing
and there is no summation over the index $j$.
Analogous relation holds for $\langle T_{00}(x, t) T_{0i}(y, t)\rangle$.
The exact result (\ref{stress}) is universal and does not depend on the frame or on the details of the
viscous and higher order terms.
In the limit of
non-relativistic  macroscopic motions, the flow velocity $v$ being much smaller than the speed of light $c$,  the  relativistic hydrodynamics equations
reduce to the non-relativistic incompressible Navier-Stokes
equations.
In this limit, the exact scaling relation (\ref{stress}) reduces in three space dimensions to the Kolmogorov four-fifths law.

These developments lead us to ask whether, as for incompressible non-relativistic fluids, statistical averages in a turbulent state of relativistic hydrodynamics exhibit a universal structure.
Note that incompressibility of the fluid implies that it is non-relativistic, since the the fluid velocity is much smaller than the speed
of sound and consequently much smaller than the speed of light.
Thus, relativistic hydrodynamics is compressible and we are interested in developing a theory for
relativistic compressible turbulence.

In the first part of this paper we propose that statistical averages in relativistic (compressible) turbulence  exhibit universal
properties.
We will consider analytically the velocity and temperature differences structure functions in the strongly intermittent (1+1)-dimensional relativistic turbulence.
We will argue that shock waves provide the main contribution to the structure functions
at the {\it bottom} of the inertial range, and
will provide evidence for scaling behavior of all structure functions of the form
\begin{equation}
\label{structure}
\langle |T(x_1)-T(x_2)|^p |\phi(x_1)-\phi(x_2)|^q\rangle \sim |x_1-x_2|^{c_{pq}},\,\,\,\,\,\,\,\,\,p,q\geq0,
\end{equation}
where $T$ and $\phi$ are the temperature and rapidity fields, and $p,q \geq0 $.  The above exponents take the values
\begin{eqnarray}
c_{pq}&=& p+q, \,\,\,\,\,\,\,\,  p+q\leq1; \label{cc1}\\
c_{pq}&=& 1,  \,\,\,\,\,\,\,\,\,\,\,\,\,\,\,\,\, p+q\geq1  \label{cc2}.
\end{eqnarray}
These hold as an intermediate asymptotic, i.e. $x_1 \rightarrow x_2$ while still satisfying the inertial range condition $|x_1 - x_2|\gg l_d$, where $l_d$ is the scale of dissipation which goes to zero in the in-viscid limit. The change of slope of $c_{pq}$ at $p+q=1$ shows that the $(1+1)$-dimensional relativistic turbulence is strongly intermittent \footnote{When $p, q$ are non-negative integers, structure functions as in (\ref{structure}) but without taking the absolute values of the arguments are well-defined. The scaling behaviors (\ref{cc1}) and (\ref{cc2}) still hold for these new functions if $p$ is even.}.

This ``phase transition'' at $p+q=1$ results from the fact that the contribution from singular objects, the shock waves, take over the contribution from the smooth component as $p+q$ increases across 1. This phenomenon is peculiar to one space dimension, in which shock waves exist as discrete objects.
These predictions can be tested numerically and potentially experimentally.
The structure functions for other quantities (such as velocity, pressure and entropy etc) can be directly derived from the above via simple change of variables and the equations of state.

The paper is organized as follows.
In section 2 we review the hydrodynamics setup and analyze the equations of motion for the relativistic viscous fluid in 1+1 dimension.
In section 3, after first reviewing the exactly solvable model of the Burgers equation to build up intuition about the dynamics of shock waves and how they contribute to correlation functions, we propose, based on a preliminary study of the shock waves in 1+1d relativistic hydrodynamics, an infinite set of the universal scaling relations. Our proposal relies on an assumption that the shocks do not cluster densely in turbulent flows. To deepen our understanding of the dynamics of the shock waves, we analyze various scattering processes involving the shocks in section 4. These include scattering of a shock against a sound wave, against a rarefaction wave, and against another shock wave. Our hope was to construct an effective interacting system consisting of shock, rarefaction, and sound waves as basic degrees of freedom, that captures much of the physics of the turbulent states, though our study falls short of this goal. We do demonstrate the stability of the shock waves along the way. This insures that, in cases without external driving force, the effect of shock waves on the short distance structure still remains as time elapses. The solution of the Riemann problem of the hydrodynamic equation plays a central role for this study of the nonlinear dynamics of shock waves. In the last section, we consider, among other things, a second type of statistical properties of the 1+1d system that is related to the events of formation of shock waves. It also contains some speculations about how to deal with the possibility of finite time blowup singularities in the 1+1d relativistic ideal hydrodynamics.

\section{The Relativistic Viscous Fluid in 1+1 Dimensions}

\subsection{Relativistic hydrodynamics}

Hydrodynamics provides a description of systems at large temporal and spatial scales, using the conserved
charges as the effective degrees of freedom. The hydrodynamics equations describe the evolution of
the charges as a series in the Knudsen number, which is effectively an expansion in gradients.

The hydrodynamics of a relativistic fluid is described by the conservation law
of the energy-momentum tensor
\begin{equation}
\label{eom}
\pd_{\mu} T^{\mu \nu} = 0 \ ,
\end{equation}
and we assume a fluid that carries no conserved charges.
The energy-momentum tensor, in a derivative expansion of the fluid dynamic variables, takes the form
\begin{equation}
\label{energymomentum}
T^{\mu \nu}=T^{\mu \nu}_{ideal} + T^{\mu \nu}_{viscous}.
\end{equation} The first, non-derivative term takes the form
\begin{equation}
\label{ideal}
T^{\mu \nu}_{ideal}=\epsilon (T) u^{\mu} u^{\nu} + p(T) \Delta^{\mu \nu} \ ,
\end{equation}  and describes an ideal fluid.
$T$, $\epsilon(T)$, and $p(T)$ are the locally-thermal-equilibrium temperature, energy density and pressure, respectively, and
\begin{equation}
\label{projection}
\Delta^{\mu \nu}= g^{\mu \nu}+ u^{\mu} u^{\nu} \ .
\end{equation}
$\Delta^{\mu \nu}$
projects onto directions orthogonal to the flow velocity $u^{\mu}$.

$T^{\mu \nu}_{viscous}$ contains a series of terms in the derivatives of $T$ and $u^{\mu}$. We will choose to define the flow velocity by the energy flow (the Landau frame). Thus, $T_{viscous}$ is constrained to be transverse to the velocity
\begin{equation}
\label{constraint}
T^{\mu \nu}_{viscous} u_{\nu} =0  \ .
\end{equation}
We can further decompose it into its trace and traceless parts. Since in $1+1$ dimension, which is the dimension we presently work in, there is no non-vanishing traceless tensor transverse to the velocity (consequently the shear viscosity term identically vanishes), to the first order of derivative, $T^{\mu \nu}_{viscous}$ takes the form
\begin{equation}
\label{viscousity}
T^{\mu \nu}_{viscous}=\zeta (T) \Delta^{\mu \nu} \nabla u \ .
\end{equation}
Here $\zeta (T)$ is conventionally called the bulk viscosity coefficient. One can also write down first order derivative terms in $T$, but all such terms can be recast into the form of the bulk viscosity by the zero-order fluid equation of motion $\pd_{\mu}T^{\mu \nu}_{ideal}=0$.

The functional dependence on $T$ of the thermodynamical state functions and the transport coefficients depends on the microscopic physics of the system one considers.

\subsection{The system in 1+1 dimensions}
For the rest of the paper, we specify ourselves to
\begin{eqnarray}
\label{equationofstates}
\epsilon &=& (2 \sigma -1 )\frac {T^{2 \sigma}} {T_0^{2 \sigma -2}}, \label{eos1} \\
p &=& \frac{T^{2 \sigma}} {T_0^{2 \sigma-2}},\label{eos2}  \\
s &=&  2 \sigma \frac{T^{2 \sigma-1}} {T_0^{2 \sigma-2}} \ , \label{eos3}
\end{eqnarray}
with $\epsilon + p = T s$.
This parameterizes a linear relation between $\epsilon$ and $p$. We will later set $T_0=1$ to avoid clustering of notations.
Such a two-dimensional non-conformal fluid may be realized by dimensionally reducing a $2 \sigma$-dimensional conformal fluid on a $2\sigma-2$-dimensional torus of volume $T_0^{-(2 \sigma-2)}$ and has a corresponding dimensionally-reduced gravity dual. It is a gravity-dilaton system on an asymptotically locally $AdS_3$ and linear dilaton background \cite{nonconformal}. The physical properties of the resulting two-dimensional fluid are expected to be analytic in $\sigma$ for $\sigma>1$. In this case, the 2d bulk viscosity is determined to be
\begin{equation}
\label{viscoeff}
\zeta = \frac{2 \sigma (\sigma-1)}{\pi (2 \sigma -1)} \frac{T^{2 \sigma -1}}{T_0^{2 \sigma-2}}\equiv \xi \frac{T^{2 \sigma -1}}{T_0^{2 \sigma-2}} \ .
\end{equation}
However, we will only study the hydrodynamics with the equation of states (\ref{eos1})-(\ref{eos3}) in this work, without resorting to the microscopic construction from the higher dimensional CFT. In particular, we will assume that the bulk viscosity $\zeta$ may be independently varied and parametrically smaller than (\ref{viscoeff}) so that we can work in the in-viscid limit, and that the fluid mean free path length may be parametrically smaller than $\frac{1}{T}$ so that the hydrodynamic description applies even for strong shocks (see section 3).

In the light-cone coordinates
\begin{equation}
\label{lightcone}
x^{\pm}=\frac{1}{\sqrt{2}}(x^0 \pm x^1) \ ,
\end{equation}
and in terms of the new variables $\phi$ and $\rho$
\begin{eqnarray}
\label{variables}
u^{\pm}&=&\frac{1}{\sqrt{2}}e^{\pm \phi},\\
T&=&e^{\rho},
\end{eqnarray}
the energy-momentum tensor reads
\begin{eqnarray}
\label{energymomentum}
T^{++}&=& e^{2 \sigma \rho+ 2 \phi} \left[\sigma-\frac{\xi}{2 \sqrt{2}} e^{-\rho} (\pd_+ e^{\phi}+\pd_- e^{-\phi} ) \right],\\
T^{--}&=& e^{2 \sigma \rho- 2 \phi} \left[\sigma-\frac{\xi}{2 \sqrt{2}} e^{-\rho} (\pd_+ e^{\phi}+\pd_- e^{-\phi} ) \right],\\
T^{+-}&=&T^{-+}=e^{2 \sigma \rho} \left[(\sigma-1)+\frac{\xi}{2 \sqrt{2}}e^{-\rho} (\pd_+ e^{\phi}+\pd_- e^{-\phi})\right].
\end{eqnarray}
The ideal approximation to the conservation laws are
\begin{equation}
\label{conserv}
\pd_+ T^{++}_{ideal} + \pd_- T^{-+}_{ideal}=\pd_+ T^{+-}_{ideal} + \pd_- T^{--}_{ideal}=0 \ .
\end{equation}
This ideal approximation is valid in the limit of slow variation and high temperature $\frac{\xi}{L T}\ll 1$, which is for instance
 the case in the inertial range of turbulence.

 The relation between the temperature and the velocity fields dictated by (\ref{conserv})
can now be solved
\begin{eqnarray}
\label{relation1}
(2\sigma-1) \pd_+ \rho &=& - \sigma \pd_+ \phi - (\sigma-1) e^{-2 \phi} \pd_- \phi, \label{rone}\\
(2\sigma-1) \pd_- \rho &=& (\sigma-1) e^{2\phi} \pd_+ \phi + \sigma \pd_- \phi \ . \label{rtwo}
\end{eqnarray} 
Equating the $\pd_-$ derivative of the $\pd_+ \rho$ equation to the $\pd_+$ derivative of the $\pd_- \rho$ equation, we arrive at
%
\begin{equation}
\label{velocity1}
2 \sigma \pd_+ \pd_- \phi + (\sigma-1) \left[\pd_+ (e^\phi \pd_+ e^\phi)-\pd_- (e^{-\phi} \pd_- e^{-\phi})\right]=0 \ .
\end{equation}
This equation does not seem to be integrable, unlike Burgers equation for non-relativistic compressible flows that
will be discussed in the next section.

In terms of the light-cone velocities, (\ref{velocity1}) takes the form
\begin{equation}
\label{velocity2}
\sigma \pd_+ \pd_- \log \frac{u^+}{u^-} + 2 (\sigma-1) \left[\pd_+ (u^+ \pd_+ u^+)-\pd_- (u^- \pd_- u^-)\right]=0.
\end{equation}
In the case of a conformal fluid, $\sigma=1$, (\ref{velocity1}) simplifies to a $(1+1)$-dimensional massless scalar field equation
\begin{equation}
\label{velocityconf}
\pd_+ \pd_- \phi =0 \ ,
\end{equation}
and \eqref{rone}, \eqref{rtwo} state that
\begin{eqnarray}
\label{relation2}
\rho &=& \rho(x^+)+\rho(x^-)=-\phi(x^+)+\phi(x^-)+c \ .
\end{eqnarray}

Small fluctuations on the background of a homogeneous flow $\phi=\phi_0+ \delta \phi$, $\rho=\rho_0+\delta \rho$
obey
\begin{eqnarray}
\label{fluctuation}
(2\sigma-1) \pd_+ \delta \rho &=& - \sigma \pd_+ \delta \phi - (\sigma-1) e^{-2 \phi_0} \pd_- \delta \phi, \label{rthree}\\
(2\sigma-1) \pd_- \delta \rho &=& (\sigma-1) e^{2\phi_0} \pd_+ \delta \phi + \sigma \pd_- \delta \phi \ . \label{rfour}
\end{eqnarray}
Their velocity and ``polarization'' are read off most easily by going to the fluid rest frame $\phi_0=0$
\begin{eqnarray}
\label{speed and polarization}
c_s&=&\frac{1}{\sqrt{2 \sigma-1}}, \label{soundspeed}\\
\frac{\delta \rho}{\delta \phi}&=& \pm c_s \label{polarization}.
\end{eqnarray}
The value of $c_s$ agrees with the expectation from the equation of states\footnote{The causality constraint $c_s\leq1$ requires that $\sigma \geq 1$. This suggests that the attempt to generate solutions to (\ref{velocity1}) perturbatively in $\sigma-1$ from solutions to the free field equation (\ref{velocityconf}) will encounter pathology. Independently, such an approach also immediately runs into infrared divergence problems.}, and the $\pm$ of the ``polarization'' apply to the right and left moving waves respectively.
Note, that unlike relativistic conformal hydrodynamics in 1+1 dimensions where the the speed of sound is
the speed of light, here for large enough $\sigma$ we can describe also non-relativistic compressible flows with velocities $c_s \leq v \ll c$.

Equations (\ref{soundspeed}) and (\ref{polarization}) together say that a small disturbance to the background travels left or right with a constant speed and a fixed profile determined by a single function ($\delta T(x^1 \pm c_s x^0)$ for example). Such a collective disturbance carries a certain amount of extra energy and extra momentum relative to the fluid at rest
\begin{eqnarray}
\delta T^{00} &=& 2 \sigma (2 \sigma-1) T^{2 \sigma-1} \delta T, \label{phonon1} \\
\delta T^{01} &=& \pm 2 \sigma \sqrt{2 \sigma-1} T^{2 \sigma-1 } \delta T, \label{phonon2} \\
\delta T^{11} &=& 2 \sigma T^{2 \sigma-1 } \delta T, \label{phonon3}
\end{eqnarray}
as well as a certain amount of extra entropy
\
\begin{eqnarray}
\delta S^0 &=& 2 \sigma (2 \sigma-1 ) T^{2 \sigma-2} \delta T, \label{phonon4} \\
\delta S^1 &=& \pm 2 \sigma \sqrt{2 \sigma-1} T^{2 \sigma-2} \delta T,   \label{phonon5}
\end{eqnarray}
all of which flip signs when the sign of $\delta T$ flips. The form of these tensors simplify when viewed against the Minkowskian metric associated to the sound-cone
\begin{equation}
\text{d} \tilde{s}^2 = - c_s^2 (\text{d} x^0)^2 + (\text{d} x^1)^2, \label{soundcone}
\end{equation}
against which $\delta T^{\mu \nu}$ becomes traceless, and $\delta T^{\mu \nu} \tilde{v}_{\nu}$ becomes proportional to $ v^\mu$ and hence null, where $v^{\mu}$ is the 2-velocity of the disturbance and $\tilde{v}_\mu = \tilde{g}_{\mu \nu} v^{\nu}$.

\section{Universal Statistics in One Dimension and Shock Waves}

Shock waves, as loci of large gradients of flow velocity and temperature, are a major source of dissipation. In their neighborhoods, entropy is produced, and kinetic energy dissipated into heat. Even in the limit in which the viscosity is taken to zero, such dissipative effects generically do not vanish, and a shock acts like a $\delta$-function source/sink for entropy/kinetic energy. This happens due to the fact that the gradients of the hydrodynamical variables often diverge as certain inverse powers of the viscosity such that the total entropy production rate and kinetic energy dissipation rate remain finite in the in-viscid limit (this is the so-called dissipative anomaly \cite{polyakov, localization}).

In $1+1$ dimension where shock waves behave as discrete structures in the in-viscid limit, the discontinuities they source in the hydrodynamical variables control the short distance behavior of an infinite set of correlation functions.


\subsection{Shocks and Structure Functions in Burgulence}

In order to build up intuition, we first review the Burgers equation, which provides an exactly solvable toy model of non-relativistic fluid motion in 1+1 dimensions.  The equation has a number of applications in a wide range of fields such as condensed matter, statistical mechanics and cosmology (for a review
of Burgulence see \cite{burgulence}).

\subsubsection{Burgers equation}

The Burgers equation reads
\begin{equation}
\label{Burgers}
\partial_t {u}+ \partial_x ({1 \over 2} u^2) = \nu \partial^2_x u + f
\end{equation} where $u$ is the velocity field, $\nu$ is the viscosity and $f(t,x)$ is an externally applied force. The time reversal symmetry $\mathcal{T}: t\mapsto -t, u \mapsto -u$ is explicitly broken for $\nu>0$, and is {\it not} restored in the in-viscid limit $\nu \searrow 0$ as a result of the dissipative anomaly. By making the Cole-Hopf transformation,
\begin{eqnarray}
\label{Cole-Hopf}
u &=& -\partial_x \phi \cr
\psi &=& e^{{1 \over {2 \nu}} \phi}
\end{eqnarray} we arrive at an imaginary time Schrodinger equation
\begin{equation}
\label{schrodinger}
- \nu \partial_t \psi = - \nu^2 \partial^2_x \psi + V \psi \ ,
\end{equation} where the potential $V$ is $V'= 2 f$ and $\nu$ plays the role of $\hbar$.

The general solution can be represented by the Feynman path integral. It suffices for us here to consider only the force-free case, for which the solution is given by convoluting the initial data with the heat kernel
\begin{equation}
\label{heat}
\psi(x,t) = \int \text{d} y {1 \over (4 \pi \nu t)^{1 \over 2}} e^{- {(x-y)^2 \over 4 \nu t}} \psi(y, 0) \ .
\end{equation} This further simplifies in the limit $\nu \searrow 0$ into:
\begin{equation}
\label{max}
\phi (x, t) = \max_{a} \{  \phi(a,0) -{ (x-a)^2 \over {2 t}} \} \ .
\end{equation}

Any local maximum of $\phi(a,0) -{ (x-a)^2 \over {2 t}}$ with respect to $a$ satisfies
\begin{equation}
\label{free}
x=a+u_0(a) t \ ,
\end{equation}which is just the trajectory of a free streaming particle.
When particle trajectories intersect, rather than freely stream across one another, the particles must collide according to the rule given by (\ref{max}),
which leads to shock formation (figure 1a). This results from local establishment of thermal equilibrium\footnote{This happens instantaneously in space and time in the in-viscid limit.}, which requires the velocity field to be a globally single-valued function of space and time coordinates.

Working with the modified, Lagrangian potential $\varphi$
\begin{equation}
\label{lagangian potential}
\varphi(a,t)\equiv t \phi(a,0)-{1 \over 2} a^2 \ ,
\end{equation} helps visualization.
Equation (\ref{max})is then equivalent to taking a Legendre-type transform of $\varphi$
\begin{equation}
\label{legendre}
t \phi(x,t) + {x^2 \over 2} = \max_{a} \{\varphi(a,t)+a x \} \ .
\end{equation} One can further replace $\varphi(a,t)$ on the right hand side of (\ref{legendre}) by its convex hull $\varphi_c(a,t)$ (defined as the intersection of all half planes containing the graph of $\varphi(a,t)$) without spoiling this identity. The shocks are easily visualized in this representation: each connected $a$-interval in which $\varphi_c(a,t)>\varphi(a,t)$ is evolved into a shock at time $t$ (figure 1b).

\begin{figure}[h!]
\vspace{-0.2cm}
\begin{center}
\scalebox{2}[2]{\includegraphics[width=0.5\textwidth]{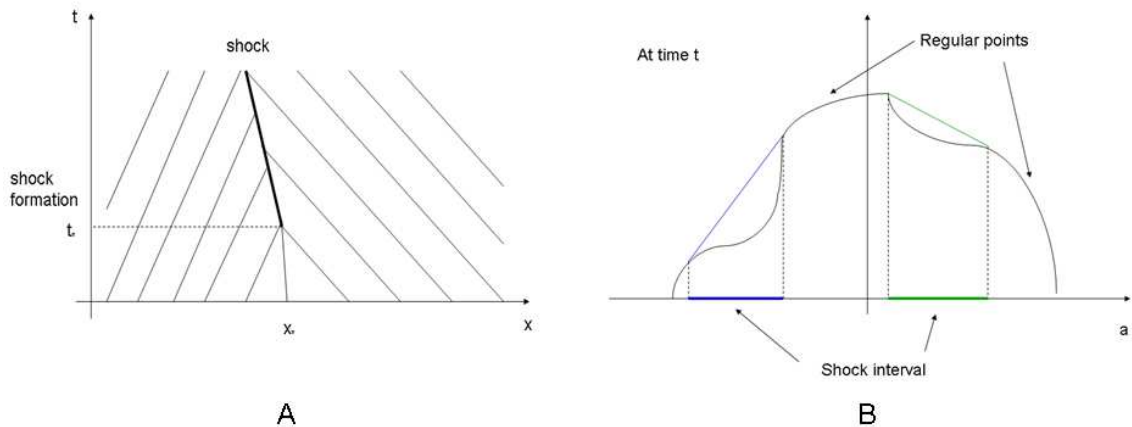}}
\end{center}
\vspace{-5.5cm}
\caption{A. A Shock forms at $(t_*, x_*)$ where free-streaming particles start to collide. Note that the inviscid limit $\nu \searrow 0$ corresponds to zooming out the space in (3.1), which is why the shock is infinitely thin in this limit, and particles stick together immediately after collision. Equivalently the mean free path of particles vanishes like $\nu$. B. $\varphi(a,t)$ vs. $\varphi_c(a,t)$ at a given time $t$. Colored intervals map to shocks.  The two graphs depict two separate processes. }
\end{figure}

\subsubsection{The Riemann problem}

The exact solution for Burgers equation (\ref{heat}) or its solution in the in-viscid limit (\ref{max}) (equivalently (\ref{legendre})) is in general not available for more complicated situations including the relativistic hydrodynamics that we will shortly study. However, if we are only concerned with the shock wave dynamics in the in-viscid limit, a more direct construction is available that determines both the propagation and interaction of shocks.

This starts by solving the Riemann problem, which is  the problem of solving hyperbolic PDE's with piece-wise constant initial conditions. Consider the initial value
\begin{eqnarray}
\label{initial}
u(x,0)&=&u_L,\,\,\,\,\,x<0 \\
u(x,0)&=&u_R,\,\,\,\,\,x>0 \ ,
\end{eqnarray} and remember that we work in the in-viscid limit $\nu \searrow 0$. As the initial value is not differentiable, to pose the Riemann problem properly requires the tool of weak solutions, which is well developed in mathematics. Here we will by-pass the mathematical discussion by resorting to the equivalent physics consideration. Instead of starting from (\ref {Burgers}) with $\nu=f=0$, we start from its integrated version, a conservation law of the form:
\begin{equation}
\label{integrated}
[{\text{d} \over \text{d} t} \int_{[a,b]} u \,\,]_t = [\,\,- \int_{\partial [a,b]} {1 \over 2} u^2 \,\,]_t= -[\,\,{1 \over 2} u(b)^2-{1 \over 2} u(a)^2\,\,]_t.
\end{equation} This is written for an arbitrary finite, fixed region of space $[a,b]$ at time $t$.  (\ref{Burgers}) is recovered where $u(x,t)$ is differentiable; otherwise (\ref{integrated}) sets the rule.

Returning to the Riemann problem. For $u_L<u_R$, fluid particles free stream from the origin. So the solution must take the form
\begin{eqnarray}
\label{rarefaction wave}
u(x,t)&=& u_L, \,\,\,\,\,x<u_L t \label{rwb1}\\
u(x,t)&=& {x \over t},\,\,\,\,u_L t < x < u_R t \label{rwb2}\\
u(x,t)&=& u_R, \,\,\,\,\,x>u_R t \label{rwb3}  .
\end{eqnarray} For $u_L>u_R$, particles free stream until they collide. Fixing the multi-valued problem after collision amounts to determining the shock velocity $s$. Without resorting to (\ref{max}) or (\ref{legendre}), one applies (\ref{integrated}) to an infinitesimal tubular neighborhood traveling with the shock
\begin{equation}
\label{matching condition}
0={\text{d} \over \text{d} t} \int^{s t+\epsilon}_{s t-\epsilon} u = s (u_R-u_L)-{1 \over 2}(u_R^2-u_L^2), \end{equation}
and reads off
\begin{equation}
\label{matching}
s={1 \over 2} (u_R+u_L).
\end{equation}

This is just the statement of momentum conservation. On the other hand, kinetic energy is decreased when particles collide.  Here we see the effect of the dissipative anomaly. The dissipative nature of particle collision forbids shock solutions (\ref{matching}) to the Riemann problem with $u_L<u_R$, even though they are mathematically valid weak solutions to (\ref{integrated}). Thus we find a complete, unique solution to the Riemann problem that is physically sensible. As expected for closed thermodynamic systems though, the existence and uniqueness of the solution hold only when it is evolved into the future direction $t>0$. If the rarefaction wave is evolved into the past, one unavoidably encounters an un-physical particle-emitting shock. Thus the rarefaction wave solution in the present context can only be made by an external agent. We will however see that they generically arise in shock wave collisions in the relativistic hydrodynamics. On the other hand, the backward evolution of the shock solution is clearly non-unique within the wedge from the origin bounded by the extensions of the free-streaming trajectories into the past.
\begin{figure}[h!]
\vspace{0.5cm}
\begin{center}
\scalebox{2}[2]{\includegraphics[width=0.5\textwidth]{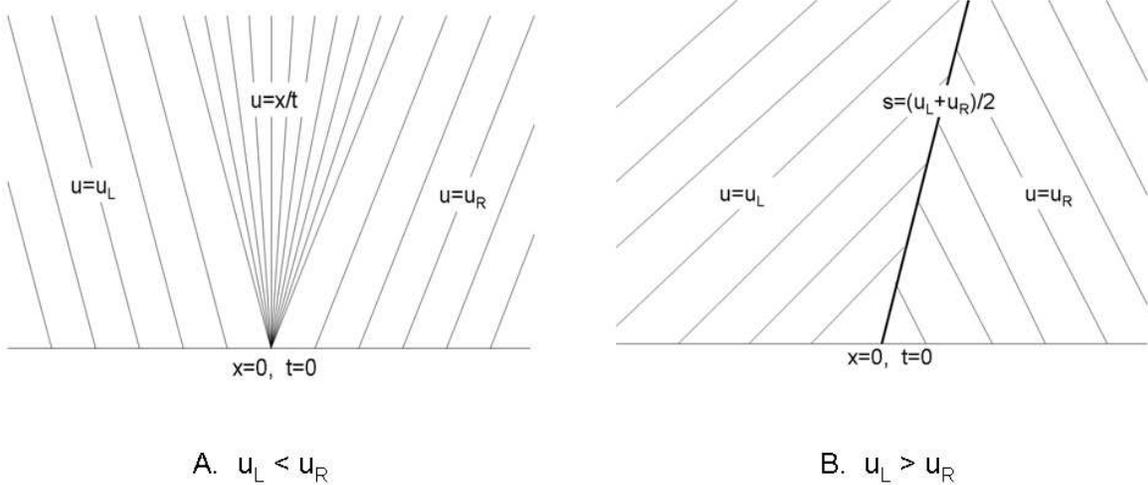}}
\end{center}
\vspace{-5.5cm}
\caption{Solutions to the Riemann problem: A. the rarefaction wave solution for $u_L<u_R$, B. the shock wave solution for $u_L>u_R$.  }
\end{figure}

This shock solution to the Riemann problem, with its uniqueness into the future and non-uniqueness into the past, provides the simple solution to the shock collision problem. Consider the initial configuration of three regions with $u_L>u_M>u_R$ separated by two shocks with $s_1 = \frac{1}{2} (u_L+u_M)$ and $s_2 = \frac{1}{2} (u_M+u_R)$. The two shocks will collide as $s_1>s_2$. On the time slice of the collision event, the configuration simplifies to the Riemann problem with $u_L>u_R$ for which we know the answer to be just a single outgoing shock with $s_3=\frac{1}{2} (u_L+u_R)$. This trivially generalizes to the case of multiple shock point-collisions, and the outcome is always a single outgoing shock with speed being the average of the velocities of the two outermost regions.

We are now ready to present a simple picture of a generic flow for the Burger equation: particles free stream across spacetime until they collide into each other to form shocks. Shocks can be created, can collide into each other and become stronger, but can not annihilate. The nature of such collisions is always many to one.  Given all these, the configuration of the shock loci consists of a set of back-ward tree-like graphs in spacetime (Figure 3.c) .

\begin{figure}[h!]
\vspace{-0.5cm}
\begin{center}
\scalebox{2}[2]{\includegraphics[width=0.5\textwidth]{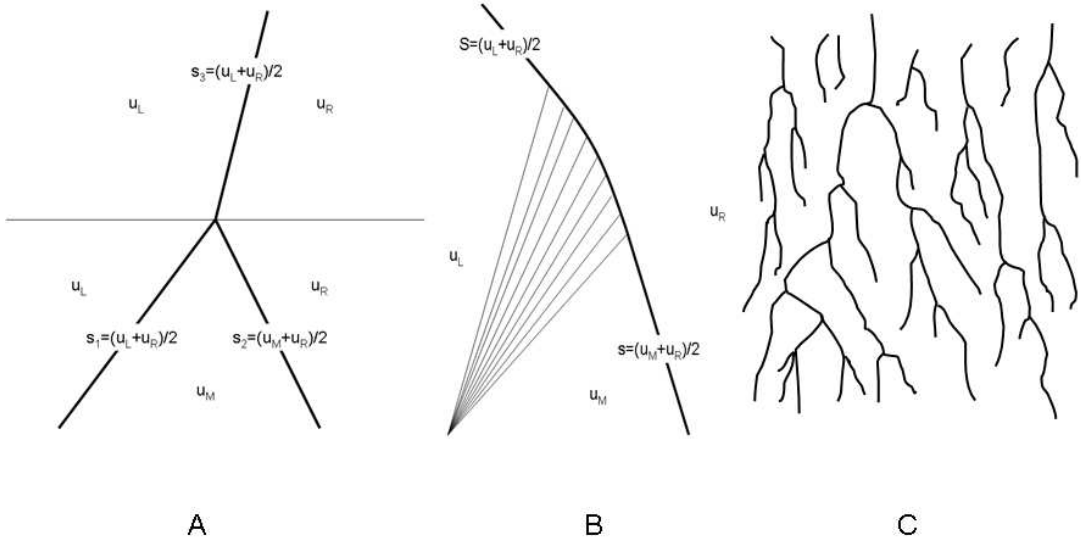}}
\end{center}
\vspace{-4cm}
\caption{Collisions between shocks and rarefaction waves: A. $2\rightarrow 1$ shock-shock collision, see the main text; B. rarefaction wave-shock collision. The curved arc in this collision is parabolic and connects to the incoming and outgoing straight sections in a $C^1$ fashion. This curve can be easily determined analytically. But since the rarefaction wave can not arise in an intermediate stage, we will not go through this simple calculation here; C. Shocks can be created, can collide into each other and become stronger, but can not annihilate. Combined with the many $\rightarrow$ one nature of the collisions, the spacetime configuration of the shock loci form tree-like structures growing into the past.}
\end{figure}

\subsubsection{Structure functions}

Consider now the velocity structure function
\begin{equation}
S_p(\Delta x, t)=\langle | u(x + \Delta x,t)-u(x,t)|^p \rangle
\end{equation}
 for an ensemble of un-forced flows with smooth and statistically homogenous initial condition. In this case, the shocks do not cluster, and has a finite number density $\rho_s(t)$ in space at any instant of time. In the limit $\Delta x \rightarrow 0$, the contribution from the smooth component is
 \begin{equation}
 P_0(\Delta x,t) \langle |u,_x(x,t)|^p\rangle_0  |\Delta x|^p \ , \nonumber
 \end{equation}
where $P_0(\Delta x,t) $ is the probability of having no shocks in the interval $[x_1,x_2]$ at time $t$, which goes to $1$ in the limit $\Delta x \rightarrow 0$. $\langle |u,_x(x,t)|^p\rangle_0$ is the average of $|u,_x(x,t)|^p$ conditional on having no shock in the neighborhood of $x$. So the contribution from the smooth component is $\propto |\Delta x|^p$ in the limit $\Delta x \rightarrow 0$.

 The contribution from the shocks, in the limit $\Delta x \rightarrow 0$, is
 \begin{equation}
 \rho_s (t)\langle|\bar{s}|^p\rangle_s(t)|\Delta x| \ , \nonumber\\
 \end{equation}
  where $\langle|\bar{s}|^p\rangle_s(t)$ is the p-th moment of the probability distribution function (PDF) of the shock amplitude (defined as $\bar{s}=u_+-u_-$) $\tilde{\rho}_s(\bar{s},t)$. Hence
\begin{equation}
S_p \sim C_p(t) |\Delta x|^p+C_p'(t) |\Delta x|.
\end{equation}
The smooth component and shocks dominate the short distance scaling of $S_p(\Delta x)$ for $0<p<1$ and $p>1$ respectively.

\medskip

\subsection{Relativistic Shocks, Profile and In-Viscid Limit}

The competition between the contributions from smooth and singular objects to structure functions generalizes to the relativistic hydrodynamics. To study this phenomenon, we first study the properties of relativistic shocks.  We focus on isolated shocks in this subsection. The dynamics of shock scattering processes will be analyzed in the next section.

\subsubsection{Ideal shocks}

An isolated shock in an otherwise homogeneous background travels with a constant profile.  In its rest frame, physics is stationary, and hence the energy and momentum fluxes are constant in space. While entropy is constantly produced at the shock front, rather than accumulates, it is carried away by the moving fluid particles to maintain global stationarity.

Away from the region close to the shock front, gradients
of velocity and temperature are small, the viscous term is negligible, and ideal hydrodynamics applies well. Constancy of the energy-momentum flux gives:
\begin{eqnarray}
f_e &=& 2 \sigma T_L^{2\sigma} \gamma_L^2 \beta_L = 2 \sigma T_R^{2\sigma} \gamma_R^2 \beta_R, \label{cont1}\\
f_p &=& T_L^{2 \sigma} \gamma_L^2 \left[(2 \sigma-1)\beta_L^2+1 \right]=T_R^{2 \sigma} \gamma_R^2 \left[(2 \sigma-1) \beta_R^2+1 \right], \label{cont2}
\end{eqnarray}
where $f_e \equiv T^{01}$ and $f_p \equiv T^{11}$ are the constant energy and momentum fluxes, L and R denote regions to the left and right side of the shock front, and the 2-velocity are parameterized by $(u^0, u^1)=\gamma (1, \beta)= \frac{1}{\sqrt{1-\beta^2}} (1,\beta)$,  and hence the matching condition between the two sides of the shock
\begin{eqnarray}
\beta_L \beta_R &=&  \frac{1}{2 \sigma-1} =c_s^2, \label{match1}\\
\frac{T_L}{T_R} &=& \left[\frac{\beta_R (1-\beta_L^2)}{\beta_L (1-\beta_R^2)} \right]^{\frac{1}{2 \sigma}}. \label{match2}
\end{eqnarray} 
Under the condition that stationary shocks exist i.e. $\sigma>1$, the fluid on one side of the shock travels with supersonic velocity and has lower temperature, the fluid on the other side travels with subsonic velocity and has higher temperature.
The entropy fluxes on the two sides therefor differ
\begin{equation}
\label{entropy}
\frac{s^1_L}{s^1_R}= \frac{s_L u^1_L}{s_R u^1_R}=\frac{2 \sigma T_L^{2 \sigma-1} \gamma_L \beta_L}{2 \sigma T_R^{2 \sigma-1} \gamma_R \beta_R}
=\left[\frac{\beta_L (1-\beta_L^2)^{\sigma-1}}{\beta_R (1-\beta_R^2)^{\sigma-1}} \right]^{\frac{1}{2 \sigma}}\neq 1,
\end{equation}
and the requirement that shock fronts create, as opposed to annihilate, entropy, dictates that the supersonic/subsonic side flows towards/away from the shock locus.

To sum up the picture, in the rest frame of the shock, upstream fluid particles come in at a supersonic velocity, they are slowed down and heated up around the shock front by viscous effects, and then go out at a subsonic velocity. The supersonic nature of the incoming flow implies that small disturbances at the shock front can only be transmitted downstream. $\beta_L \approx \beta_R \approx \pm c_s$ in the limit of small discontinuities because, in this limit, the shock wave can be approximated by a small disturbance that necessarily travels with the speed of sound relative to the background fluid.

\subsubsection{Viscosity and the in-Viscid limit}

Close to the transition region around the shock front, the naive discontinuities are resolved by the viscosity effects. The bulk viscosity term contributes the derivative terms to the energy-momentum flux,
\begin{eqnarray}
\label{contvis}
f_e&=& (2 \sigma \gamma^2 \beta) T^{2 \sigma}- (\xi \gamma^5 \beta \, \beta') T^{2 \sigma-1}\label{contvis1},\\
f_p&=& \gamma^2 \left[(2 \sigma-1) \beta^2+1 \right] T^{2 \sigma} - (\xi \gamma^5 \,\beta') T^{2\sigma-1},\label{contvis2}
\end{eqnarray}
where $\beta'$ denotes $\frac{d \beta}{d x}$ and $\gamma = \frac{1}{\sqrt{1 -\beta^2}}$.
The two constant solutions, i.e. the two roots $\beta_{\pm}$ of
\begin{equation}
\label{fixed points1}
\beta^2 - \frac{2 \sigma}{2 \sigma-1} \left(\frac{f_p}{f_e}\right) \beta + \frac{1}{2 \sigma-1} =0
\end{equation}
and the associated
\begin{equation}
\label{fixed points2}
T_{\pm}= \left(\frac{f_e}{2 \sigma} \frac{(1-\beta_{\pm}^2)}{ \beta_{\pm}}\right)^{\frac{1}{2 \sigma}}
\end{equation}
are the asymptotic values of the velocity and temperature fields to the far left and far right of the shock front. The smoothed shock solution to (\ref{contvis1}) and (\ref{contvis2}) interpolates between them.

Such interpolating solutions generically exist, as we now show. Regarding (\ref{contvis1}) and (\ref{contvis2}) as two coupled linear equations for $T^{2 \sigma}$ and $T^{2 \sigma-1}$, we find
\begin{eqnarray}
T^{2 \sigma} &=& \frac{1}{2 \sigma-1} (\frac{f_e}{\beta}-f_p)\equiv A(\beta; f_e,\,f_p;\sigma) \label{T1},\\
T^{2 \sigma-1}&=& \frac{\left[(2 \sigma-1) \beta^2+1 \right]f_e - 2 \sigma \beta f_p }{(2 \sigma -1) \xi \gamma^3 \beta \beta'} \equiv \frac{B(\beta; f_e,\,f_p; \sigma)}{\beta'},\label{T2}
\end{eqnarray}
where two functions $A(\beta; f_e,\,f_p;\sigma)$ and $B(\beta; f_e,\,f_p; \sigma)$ are introduced to avoid lengthy equations in the following. Note that
\begin{equation}
\label{zero}
A|_{\beta=f_e/f_p}= B|_{\beta=\beta_{\pm}}=0.
\end{equation}
In addition to the requirement\footnote{From now on, we fix our space orientation so that $f_e>0$. So the L and R sides are the incoming upstream and outgoing downstream sides of the shock.}
\begin{equation}
\label{condition1}
0<\beta_-<\beta_+<1,
\end{equation}
$f_e$, $f_p$ and $\sigma$ are required by the positivity of temperature (\ref{T1}) to obey
\begin{equation}
\label{condition}
\frac{f_e}{f_p}>\beta_+,
\end{equation}
such that $T^{2 \sigma}=A(\beta)$ is positive definite for $\beta_-\leq \beta \leq \beta_+$.  Since $B(\beta)\leq0$ in the same region, (\ref{T2}) implies that $\beta' \leq 0$, which is consistent with the physical intuition that viscous effects monotonically slow down the motion of the fluid particles.

Equations (\ref{T1}) and (\ref{T2}) together give:
\begin{equation}
A(\beta)^{2 \sigma-1} = (\frac{B(\beta)}{\beta'})^{2 \sigma}, \label{T3}
\end{equation}
which always has solutions interpolating $\beta_{\pm}$ between $x=\pm \infty$.  One may interpret this ODE as describing a point particle moving along the $\beta$-coordinate, along which there is a potential
\begin{equation}
\label{potential}
V(\beta)\equiv -\frac{1}{2} \frac{B(\beta)^2}{A(\beta)^{2-1/\sigma}},
\end{equation}
which is smooth and negative for $\beta_-<\beta<\beta_+$ and vanishes at the two local maxima at $\beta=\beta_{\pm}$. Interpreting $x$ as the time direction for this particle, (\ref{T3}) is the statement of vanishing energy
\begin{equation}
\label{energy}
E=\frac{1}{2} \beta'^2 + V(\beta)=0.
\end{equation}
Because of energy conservation, any trajectory starting from $\beta=\beta_0$ ($\beta_-<\beta_0<\beta_+$) at the initial time $x=x_0$ with zero-total energy (\ref{energy}) approaches $\beta=\beta_{\pm}$ in the past and the future, and it does so asymptotically (as opposed to in finite time) as a result of the quadratic behavior of the potential near these two points. These are the smooth shocks we look for:
\begin{eqnarray}
\label{shock}
-x+x_0 &=& \int^{\beta}_{\beta_0} \text{d}\beta  \frac{A(\beta)^{1-\frac{1}{2 \sigma}}}{|B(\beta)|}  \label{s1} \nonumber \\
&=& \xi f_e^{- \frac{1}{2 \sigma}} \int^{\beta}_{\beta_0} \text{d}\beta \{ \gamma^3 \beta^{\frac{1}{2 \sigma}} \frac{[\frac{1}{2 \sigma-1} (1-\frac{f_p}{f_e} \beta)]^{1-\frac{1}{2 \sigma}}}{(\beta_+-\beta)(\beta-\beta_-)}\}  \label{s2} \nonumber\\
&\equiv& \xi f_e^{-\frac{1}{2 \sigma}} F(\beta, \beta_0; \sigma, \frac{f_p}{f_e})  \ .
\label{s3}
\end{eqnarray}
Modulo spatial translations and reflection, physically inequivalent shocks are parameterized by $|f_e|$, $f_p$ and $\sigma$. As (\ref{s3}) shows, the shape of the shock profile is controlled by  $\frac{f_p}{|f_e|}$ and $\sigma$, and its overall width is proportional to $\xi (|f_e| T_0^{2 \sigma-2})^{-\frac{1}{2 \sigma}}$, where we restore the dimension with $T_0$.

In the inviscid limit $\xi \searrow 0$, the width of the transition region goes to zero, and the internal structure of the shocks are effectively replaced by a discontinuity satisfying the matching conditions (\ref{match1}) and (\ref{match2}). Furthermore, the divergence of the entropy-current $0<\pd_{\mu} S^{\mu}=\pd_{\mu} (s u^{\mu}) = \pd_1 (s u^1) $ approaches a $\delta$-function supported at the discontinuity, and the shock front acts like a $\delta$-function source of entropy that causes the entropy flux discontinuity (\ref{entropy}), computed based on the ideal hydrodynamical approximation. Here we arrive at this effective singular shock description by sending the bulk viscosity to zero, and we expect this effective description of shocks in the in-viscid limit is generically valid regardless of the type of regulation one uses\footnote{For example, one may look for smooth shock solutions with additional higher derivative terms in $T^{\mu \nu}_{viscous}$ other than the bulk viscosity term. The apparent universal validity of the ideal analysis of the previous section suggests that we should reproduce the same result in the in-viscid limit after sending all the higher transport coefficients to zero.}.

\subsection{Universal Statistics of Relativistic Turbulence due to Shocks}

Consider now an ensemble of relativistic flows that, under external driving force, reached a statistically stationary state \footnote{Due to the entropy production at the shock fronts, to maintain such a stationary state, a mechanism is needed to release the created entropy into the environment. Otherwise the average temperature of the system increases monotonically. This, a priori, does not affect the discussion of the structure functions as they concern temperature and velocity differences between two points. }. Assume that there is a finite density of shocks in the resulting stationary state, in particular we exclude the situations where the shocks cluster densely. In the limit $|x-y| \rightarrow 0$, to any structure function of the form
\begin{equation}
\label{structure2}
\langle|\phi(x,t)-\phi(y,t)|^p |\rho(x,t)-\rho(y,t)|^q\rangle,
\end{equation}
the contribution from shocks scale as $|x-y|$ and the contribution from the smoother component of the flows scale as $|x-y|^{p+q}$. The former scaling arises because shocks source discontinuities in $\rho$ and $\phi$, and the latter arises by Taylor expanding the smoother components of $\rho(x,t)$ and $\phi(x,t)$ to the first order in $x$. One may check that, in the strict in-viscid limit where the weak discontinuities (i.e. locations where the first derivatives of the variables $\rho$ and $\phi$ jump, such as the edges of the centered rarefaction waves) remain localized, the presence of such weak discontinuities does not introduce scaling behavior that dominates $|x-y|^{p+q}$. As $p+q$ increases across $1$, we observe the transition behavior (\ref{cc1}) and (\ref{cc2}).

We note that the set of universal scalings (\ref{structure}) apply to a different range of length scales from (\ref{stress}) does. Both apply to lengths larger than the scale of dissipation, however on the upper end (\ref{structure}) breaks down below the mean distance scale between shocks, and (\ref{stress}) breaks down at the large scale where boundary effects set in. So generically, we expect (\ref{stress}) to apply to a larger range of scales. Near the bottom of the inertial range where both sets of scalings apply, they are consistent with each other
\begin{eqnarray}
\label{consistent}
& &\langle |T_{01}(x,t)-T_{01}(y,t)||T_{11}(x,t)-T_{11}(y,t)| \rangle \cr
&=& |\langle T_{01}(x,t)T_{11}(x,t)\rangle | + | \langle T_{01}(y,t)T_{11}(y,t) \rangle | + | \langle T_{01}(x,t)T_{11}(y,t) \rangle | \cr
& & + | \langle T_{01}(y,t)T_{11}(x,t) \rangle | \cr
&=& 2 \epsilon |x-y|.
\end{eqnarray}

\medskip

We also note that it is an issue of dynamics whether the shocks cluster densely so that the scaling relation (\ref{structure}) breaks down. In the context of Burgulence, it is known, via numerical as well as analytic studies, that the distribution of shocks depends also on the spectral property of the ensemble of initial conditions in addition to that of the external driving force, and that both clustering and non-clustering behavior may happen \cite{sinai} \cite{saf}. Our knowledge of the relativistic hydrodynamics is at present far from enough for resolving this issue. It is conceivable that, among other things, the presence of propagating sound waves in the relativistic case leads to physics qualitatively different from that of Burgers, and alleviates the dependence on the properties of the initial states ensemble.

\bigskip

\section{Interacting Dynamics of Relativistic Shocks}

The generic shocks are non-isolated and non-stationary. In constant interaction with their environment, their amplitudes, profiles, and velocities change over time. Furthermore, at non-vanishing viscosity where the shock front is smoothed out by diffusion, the definition of a shock locus depends on convention.  It appears to be a difficult task to analyze the problem of nonlinear shock evolution.

Here we are interested in understanding the interacting shock physics in the {\it in-viscid} limit, where based on simple physical assumptions, large amount of quantitative information can be extracted about their scattering.
Concretely, we analyze the scattering problem between shock wave and sound wave, which is accomplished by a perturbative analysis of the junction condition across the shock front; we also solve the scattering problem between shock wave and shock wave, and shock wave and rarefaction wave, which is accomplished by constructing entropy-nondecreasing weak solutions from the complete solution of the Riemann problem associated to the ideal relativistic hydrodynamics.

We set out this analysis in the hope of constructing an effective interacting model consisting of shocks, rarefaction waves, and sound waves as basic ingredients, that captures much of the nonlinear physics of turbulence. However, the results collected here fall short of this goal, which will be left for future work.

\subsection{Shock-Sound Scattering and Perturbative Shock Stability}

In this subsection we probe the physics of the shock front by its scattering with sound waves. We consider three situations. First, a small disturbance may be sent towards the shock front from upstream. No reflected sound wave can arise in this case since the upstream fluid flows supersonically towards the shock front, only transmitted sound wave emerges downstream. Second, a small downstream disturbance traveling against the subsonic outgoing flow towards the shock will eventually reach the shock. No transmitted wave can emerge in this case, only the reflected wave arises. And finally, disturbance at the shock front due to thermal fluctuations may cause spontaneous emission of sound waves towards downstream. For such decay to materialize, there needs to exist classical solutions with only outgoing sound waves and no incoming ones. We will perform the analysis in the in-viscid limit and in the probe approximation, and a by-product of our result is the establishment of the perturbative stability of shock front against spontaneous emission of sound waves.

\medskip

Away from the shock front, each small disturbance propagates relative to its background flow according to (\ref{soundspeed}) and (\ref{polarization}) and is determined by an arbitrary profile function. At the shock front, whose thickness in the in-viscid limit is negligible compared to the wavelengths of the disturbances, the shock front sees the disturbances as providing it with a time-dependent environment $\beta_{L,R}(x^0, x^1_s)$ and $T_{L,R}(x^0,x^1_s)$. Thus the junction condition for the disturbances amounts to imposing the constant-flux condition (\ref{cont1}) and (\ref{cont2}) in the instantaneous comoving frame of the shock
\begin{equation}
\label{junction}
(T^{\mu \nu}_L-T^{\mu \nu}_R + \delta T^{\mu \nu}_L- \delta T_R^{\mu \nu}) \Delta^s_{\nu \rho}=0 \ .
\end{equation}
Here
\begin{equation}
\label{shockprojector}
\Delta^s_{\nu \rho}=g_{\nu \rho}+u^s_{\nu} u^s_{\rho} \ ,
\end{equation}
where $u^s_\mu$ is the 2-velocity of the shock front, projects onto the spatial direction in the comoving frame of the shock.
$\delta T^{\mu \nu}_{L,R}$ depends on the values of the disturbances at the shock locus.  (\ref{junction}) generically provides two constraints. In the case of transmission or reflection, they generically determine the amplitudes of outgoing sound wave and the shock front oscillation in terms of the amplitude of the incoming disturbance. In the absence of an incoming wave, they generically determine that the emission is zero. If there is a further degeneracy in (\ref{junction}) so that less than two constraints occur in (\ref{junction}), the transmission and reflection amplitudes would be under-determined as a result of non-vanishing effect of spontaneous emission, and the shock would be unstable.

Starting with the case of transmission. (\ref{junction}) gives
\begin{eqnarray}
0&=&(T^{00}_L-T^{00}_R+\delta T^{00}_L-\delta T^{00}_R) \beta_s  - (\delta T^{01}_L-\delta T^{01}_R), \label{c1}\\
0&=&(\delta T_L^{10}-\delta T_R^{10}) \beta_s - (\delta T_L^{11}-\delta T_R^{11}). \label{c2}
\end{eqnarray}
$\delta T^{\mu \nu}_{L,R}$ are the energy-momentum tensors for the incoming and outgoing disturbances, each of them is a function of the amplitude of the corresponding disturbance.
\begin{eqnarray}
\label{energymomentum3}
\delta T^{\mu \nu}_{\pm} &=& \delta_{\pm}\{T^{2 \sigma} (2 \sigma u^\mu u^\nu+g^{\mu \nu}) \} \cr
&=&2 \sigma T^{2 \sigma}\{(2 \sigma u^\mu u^\nu +g^{\mu \nu}) (\frac{\delta T}{T}) + (\delta u^\mu u^\nu + u^\mu \delta u^\nu) \}|_{\pm} \cr
&=&2 \sigma T^{2 \sigma} \{(\frac{\delta T} {T}) (2 \sigma \gamma^2  \begin{bmatrix} 1&\beta\\ \beta&\beta^2 \end{bmatrix} + \begin{bmatrix} -1&0\\ 0&1 \end{bmatrix} ) + \gamma^4 \delta \beta \begin{bmatrix} 2 \beta&1+\beta^2\\ 1+\beta^2&2 \beta \end{bmatrix} \}_{\pm} \cr
&=&2 \sigma T^{2 \sigma-1} \delta T \{2 \sigma \gamma^2  \begin{bmatrix} 1&\beta\\ \beta&\beta^2 \end{bmatrix} +  \begin{bmatrix} -1&0\\ 0&1 \end{bmatrix}  \pm \sqrt{2 \sigma-1} \gamma^2 \begin{bmatrix} 2 \beta&1+\beta^2\\ 1+\beta^2&2 \beta \end{bmatrix} \}  \cr
&=&2 \sigma T^{2 \sigma} \gamma^2 \delta \beta  \{\pm 2 \sigma \gamma^2 \begin{bmatrix} 1&\beta\\ \beta&\beta^2 \end{bmatrix} \pm \begin{bmatrix} -1&0\\ 0&1 \end{bmatrix}  + \sqrt{2 \sigma-1} \gamma^2  \begin{bmatrix} 2 \beta&1+\beta^2\\ 1+\beta^2&2 \beta \end{bmatrix} \} \cr
&=& (\text{D}_T T_{\pm}^{\mu \nu}) \delta T = (\text{D}_\beta T_{\pm}^{\mu \nu}) \delta \beta \label{deriv}
\end{eqnarray}
From the third line to the forth and fifth lines of (\ref{deriv}), we plugged in the relation (\ref{polarization}) in the form of
\begin{equation}
\label{polarization2}
\frac{\delta T}{\delta \beta}=\pm \frac{\gamma^2 T}{\sqrt{2 \sigma-1}}
\end{equation}
The last line of (\ref{deriv}) defines the derivatives $\text{D}_T T_{\pm}^{\mu \nu}$and $\text{D}_\beta T_{\pm}^{\mu \nu}$. We choose the $+$ sign in the case of transmission since both waves are right-moving.  It is clear that there is no degeneracy between (\ref{c1}) and (\ref{c2}) at least in the probe approximation where $\delta T_{L,R}$ are small, for which (\ref{c1}) and (\ref{c2}) linearize to
\begin{eqnarray}
0&=&(T^{00}_L-T^{00}_R) \beta_s  - (\delta T^{01}_{L,+}-\delta T^{01}_{R,+}), \label{d1}\\
0&=&\delta T_{L,+}^{11}-\delta T_{R,+}^{11}. \label{d2}
\end{eqnarray}
and we read off the amplitudes for the transmitted wave and the shock excitation
\begin{eqnarray}
\frac{\delta T^{out}}{\delta T^{in}} &=& \frac{(\text{D}_T T_+^{11})_L}{(\text{D}_T T_+^{11})_R}=\frac{T_R}{T_L}, \label{transm1} \\
\beta_s &=& \frac{(\delta T_+^{01})_L-(\delta T_+^{01})_R}{T^{00}_{L} -T^{00}_R} \\
&=& \frac{[(\text{D}_T T_+^{01})_L (\text{D}_T T_+^{11})_R -(\text{D}_T T_+^{01})_R (\text{D}_T T_+^{11})_L] \delta T^{in}}{(\text{D}_T T_+^{11})_R (T_L^{00}-T_R^{00})}. \label{oscillation}
\end{eqnarray}
We note two points here. First, concerning the evolution of the disturbance, the transmission coefficient
in terms of $\frac{\delta \rho^{out}}{\delta \rho^{in}}=\frac{\delta \phi^{out}}{\delta \phi^{in}}$ is a constant (independent of the wavelengths of the incoming disturbance) and equals exactly to $1$. The sound wave incoming from upstream passes through the shock front without any modification.
Second, concerning the back-reaction on the shock, the direction of the ``kick'' the shock receives from the disturbance depends on whether the disturbance is a temperature surplus or deficit relative to the background. If the shock is hit by a bump of temperature surplus $\delta T>0$, it experiences a ``positive kick''and acquires a downstream motion $\beta_s>0$; and if hit by a bump of temperature deficit $\delta T<0$, it feels a ``negative kick'' and moves upstream $\beta_s>0$.
The reactive motion of the shock ceases after the disturbance passes by, and total displacement of the shock front is proportional to the energy surplus/deficit the sound wave carries.

Similar analysis can be done for the cases of reflection and potential spontaneous emission. In the reflection case, (\ref{c1}) and (\ref{c2}) linearize in the probe approximation to
\begin{eqnarray}
0&=&(T^{00}_L-T^{00}_R) \beta_s  + (\delta T^{01}_{R,+}+\delta T^{01}_{R,-}), \label{e1}\\
0&=&-\delta T_{R,+}^{11}-\delta T_{R,-}^{11}. \label{e2}
\end{eqnarray}
From these we read off the amplitudes for the reflected wave and the shock excitation
\begin{eqnarray}
\frac{\delta \rho^{out}}{\delta \rho^{in}}=\frac{\delta T^{out}}{\delta T^{in}} &=& -\frac{(\text{D}_T T_-^{11})_R}{(\text{D}_T T_+^{11})_R}=-\frac{\sigma f_p-\sqrt{2 \sigma-1} f_e}{\sigma f_p+\sqrt{2 \sigma-1} f_e}<0, \label{transm2} \\
\beta_s &=& -\frac{(\delta T_-^{01})_R+(\delta T_+^{01})_R}{T^{00}_{L} -T^{00}_R} \\
&=& -\frac{[(\text{D}_T T_-^{01})_R (\text{D}_T T_+^{11})_R -(\text{D}_T T_+^{01})_R (\text{D}_T T_-^{11})_R] \delta T^{in}}{(\text{D}_T T_+^{11})_R (T_L^{00}-T_R^{00})}. \label{oscillation}
\end{eqnarray}
The reflection amplitude $\frac{\delta T^{out}}{\delta T^{in}} $ is negative, its magnitude strictly less than $1$, and vanishes when $\beta_R \nearrow c_s$, a reflection of the fact that in this limit the incoming wave ceases to be able to reach the shock front. The ratio $\frac{\beta_s}{\delta T^{in}}$, as it should, also vanishes as $\beta_R \nearrow c_s$, and is otherwise negative.

It should be clear by now that no purely outgoing sound wave solution exists, as $(\text{D}_T T_+^{11})_R>0$. Thus the shock is perturbatively stable against spontaneous emission of sound waves\footnote{With the issue of perturbative, radiative stability now closed, one may want to go back to solve again the shock-sound scattering problem ((\ref{junction})) in a different regime---the soft shock regime. It is interesting because in this regime both the shock and the disturbance may be approximately described by sound waves, and one may learn about sound-sound scattering by solving (\ref{c1}) and (\ref{c2}).}.

\subsection{Nonlinear Shock Scattering}

A generic flow consists of a smooth component of sound waves with large, variable amplitudes as well as various shock waves of variable amplitudes, which furthermore transform into each other under time evolution. To solve their interaction dynamics in generality requires dealing with non-linear PDE's. Here we will limit ourselves to simple cases of interaction between shock fronts and between shock front and simple waves. Solutions in this context may be found based on solving the Riemann problem. This generalizes the corresponding discussion in subsection 3.1.2 in the case of Burgers, and the gist of the approach is to look for entropy non-decreasing weak solutions of the {\it ideal} hydrodynamic equations. We note that, in contrast to Burgers where this approach may be rigorously justified by taking the $\nu \rightarrow 0$ limit of the exact solution, it is our working assumption in the present context that this approach reproduces correctly the in-viscid limit of the interacting dynamics of the shock waves in the viscous hydro. The validity of this approach is supported by the reasonableness of the solutions we derive.

\subsubsection{Solution to the Riemann problem}

Here we present the solution to the Riemann problem that generalizes the corresponding discussion in the Burgers case.  Assume the initial condition at $t=0$,
\begin{eqnarray}
\label{riemann}
(\phi, \rho)  &=& (\phi_L, \rho_L)  \,\,\,\,\,\,\,\,\,\, x<0 \label{left} \\
(\phi, \rho)  &=& (\phi_R, \rho_R)  \,\,\,\,\,\,\,\,\,\, x>0. \label{right}
\end{eqnarray}
Physics dictates the existence and uniqueness of the sensible future evolution. We proceed to construct it by gluing together sensible solutions of simple types \cite{PDE}.

First we attempt to interpolate the future evolution of (\ref{left}) and (\ref{right}) by a shock wave. This is sensible only if the discontinuity at the origin describes a compressed region. In this case, i.e.
\begin{equation}
\label{riemann1}
\phi_L> \phi_R,
\end{equation}
we solve the boost of (\ref{match1})
\begin{equation}
\label{match3}
\text{tanh}(\phi_L-\phi_0)\, \text{tanh}(\phi_R-\phi_0)=c_s^2,
\end{equation}
and find
\begin{equation}
\label{match4}
\text{e}^{2 \phi_0-\phi_L-\phi_R}=\frac{(1+c_s^2) \text{cosh} (\phi_L-\phi_R)\pm \sqrt{(1+c_s^2)^2 \text{cosh}^2(\phi_L-\phi_R)-(1-c_s^2)^2} }{1-c_s^2}.
\end{equation}
Here $\phi_0$ is the rapidity of the shock front, (\ref{match3}) is a rewriting of (\ref{match1}) in terms of the rapidity variables, and $+$ in (\ref{match4}) describes the case of the left/right side being the downstream/uptream side, and $-$ describes the other case. The relation between $\rho_L$ and $\rho_R$ is constrained by (\ref{match2}), which now takes the form
\begin{eqnarray}
\rho_R-\rho_L &=& \text{log} \frac{T_R}{T_L} = \frac{1}{2 \sigma} \text{log} \frac{\beta_L (1-\beta_R^2)}{\beta_R (1-\beta_L^2)}\cr
&=& -\frac{1}{2 \sigma} \text{log}[ \text{sinh} (2 \phi_R-2 \phi_0)]+ \frac{1}{2 \sigma} \text{log}[ {\text{sinh}({2 \phi_L-2 \phi_0})}]\cr
&=& -\frac{1}{2 \sigma} \text{log} \frac{[\sqrt{\sigma^2 \text{cosh}^2 (\phi_R-\phi_L)-(\sigma-1)^2}\mp\sigma \text{sinh} (\phi_R-\phi_L)]^2}{2 \sigma-1}, \label{shockcurve}
\end{eqnarray}
where we used (\ref{match4}) from the second to the third line above. (\ref{shockcurve}), together with (\ref{riemann1}), defines the two semi-infinite shock curves through the point $(\phi_L,\rho_L)$. The compressive condition (\ref{riemann1}) is equivalent to the condition of positive entropy production. In the soft-shock limit $\phi_L-\phi_R \searrow 0$, (\ref{shockcurve}) simplifies into the expected form
\begin{equation}
\rho_R-\rho_L \sim \pm c_s (\phi_R- \phi_L). \label{softshock}
\end{equation}
(\ref{shockcurve}) also simplifies in the strong shock regime $\phi_L \gg \phi_R $
\
\begin{equation}
\rho_R-\rho_L \sim \pm \frac{1}{\sigma} (\phi_R- \phi_L) \mp \frac{1}{\sigma} \text{log} \frac{\sigma^2}{2 \sigma-1} + o (e^{- 2(\phi_L-\phi_R)}). \label{strongshock}
\end{equation}
This latter fact implies an upper bound on the strength of the rarefaction wave generated in the shock-shock catch-up scattering processes.

We next find special solutions to the Riemann problem in the form of simple waves, which, by an abuse of notation, are defined by the ansatz
%
\begin{equation}
\rho (x^+, x^-) = \rho(\phi (x^+,x^-)). \label{simple}
\end{equation}
These are generalizations of the sound waves to the nonlinear regime, retaining the feature of the single-valued relation between $\rho$ and $\phi$. Feed this simple wave ansatz (\ref{simple})
into (\ref{rone}) and (\ref{rtwo}), one arrives at
\begin{eqnarray}
(2 \sigma-1) \frac {\text{d} \rho} {\text{d} \phi} \partial_+ \phi &=& -\sigma \partial_+ \phi - (\sigma-1) e^{-2 \phi} \partial_- \phi,  \label{s1}\\
(2 \sigma-1) \frac {\text{d} \rho} {\text{d} \phi} \partial_- \phi &=&  \sigma \partial_- \phi + (\sigma-1) e^{2 \phi} \partial_+ \phi.  \label{s2}
\end{eqnarray}
Solving (\ref{s1})(\ref{s2}) gives
\begin{eqnarray}
&&\frac{\text{d}\rho}{\text{d} \phi} = \pm c_s, \label{s3}\\
&&\partial_- \phi + c_{\pm} e^{2 \phi} \partial_+ \phi = 0. \label{s4}
\end{eqnarray}
Here $c_s=\frac{1}{\sqrt{2 \sigma-1}}$ is as in (\ref{soundspeed}) the speed of sound, and $c_{\pm} = \frac{\sigma \pm \sqrt{2 \sigma-1}}{\sigma-1}$ is the light-cone speed for right/left traveling sound. The $\pm$ signs in (\ref{s1}) and (\ref{s2}) are correlated, and describe right and left moving simple waves. (\ref{s3}) generalizes (\ref{polarization}) and determines $\rho$ in terms of $\phi$. (\ref{s4}) may be rewritten in the form of Burgers equation
\begin{equation}
\label{s5}
\partial_- e^{2 \phi}+ c_{\pm} e^{2 \phi} \partial_+ e^{2 \phi}=0,
\end{equation}
and this suggests the analogue of the free-streaming ansatz (\ref{rarefaction wave}) in the case
\begin{equation}
\label{rarefy}
\phi_L<\phi_R.
\end{equation}
Indeed, one easily sees that
\begin{eqnarray}
\label{rarefaction wave 3}
\phi(x,t)&=& \phi_L, \,\,\,\,\,\,\,\,\,\,\,\,\,\,\,\,\,\,\,\,\,\,\,x<u_{L,\pm} t, \label{rw1}\\
\phi(x,t)&=& \frac{1}{2} \text{log} \frac{x^+}{c_{\pm} x^-},\,\,\,\,u_{L,\pm} t < x < u_{R,\pm} t, \label{rw2}\\
\phi(x,t)&=& \phi_R, \,\,\,\,\,\,\,\,\,\,\,\,\,\,\,\,\,\,\,\,\,\,\,x>u_{R,\pm} t. \label{rw3}
\end{eqnarray}
solves (\ref{s5}) under the condition (\ref{rarefy}). The values of $u_{L,\pm}$ and $u_{R,\pm}$ are set by matching the values of $\phi$ across the rarefaction wave boundaries
\begin{eqnarray}
\frac{1+u_{L,\pm}}{1-u_{L,\pm}}&=&c_{\pm} e^{2 \phi_L}, \label{uL}\\
\frac{1+u_{R,\pm}}{1-u_{R,\pm}}&=&c_{\pm} e^{2 \phi_R}. \label{uR}
\end{eqnarray}
Hence across the boundaries of the centered rarefaction waves, the first derivatives of the fields $\phi$ and $\rho$ are discontinuous. Such boundary of the rarefaction waves propagate relative to the background fluid at the speed of the sound, which is necessarily so since one may smooth it out into small bumps that propagate as sound disturbances. These weak discontinuities behave differently from shock fronts with a positive viscosity turned. While shock fronts remain localized within a region of width set by the viscosity, the weak discontinuities broaden out. The above solutions are illustrated in Figure 4 (the captions are self-explanatory).
(\ref{s3}) together with the condition (\ref{rarefy}) determine the two semi-infinite rarefaction curves emitted from the point $(\phi_L, \rho_L)$.
\begin{figure}[h!]
\vspace{-0.35cm}
\begin{center}
\scalebox{2}[2]{\includegraphics[width=0.5\textwidth]{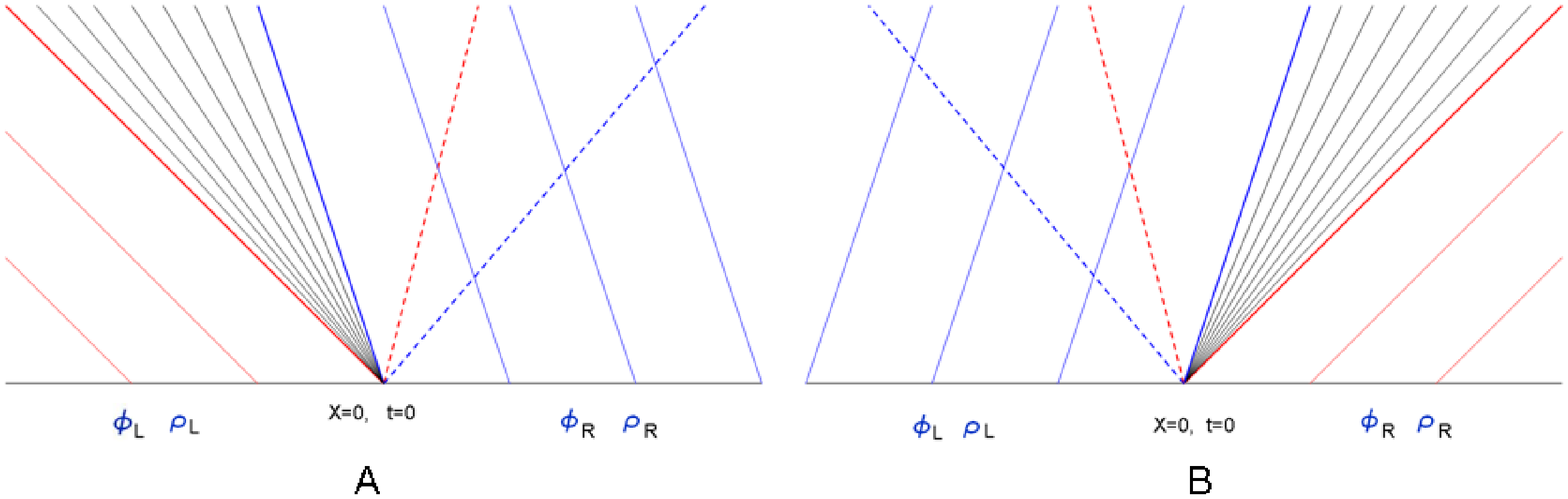}}
\end{center}
\vspace{-6.0 cm}
\caption{Rarefaction wave solutions to the Riemann problem with $\phi_L<\phi_R$:\,\,\,\,\,\,\,\,\,\,\,\,\,\,\,\,\,\,\,\,\,\,\,\,\,\,\,\,\,\,\,\,\,\,\,\,
A. The left-moving rarefaction wave solution for $\rho_R-\rho_L=-c_s (\phi_R-\phi_L)$. This corresponds to choosing the $-$ sign in (4.28)$-$(4.34)
. 
The parallel straight lines in the figure to the two sides of the central fan-like region are the left-moving characteristics of (4.29) 
in those regions, i.e. they are the space-time trajectories of left-moving virtual sound signals relative to the fluids on the two sides of the discontinuity.
The red/blue (dashed)lines from the origin are the future sound cones from the origin relative to the fluids on the left/right side. Outside of the union of the future sound-cones of the discontinuity, the fluids maintain its constant states without knowledge of the discontinuity. In the region between, the fluid behaves according to the conservation laws, which, for the present case i.e. $\rho_R-\rho_L=-c_s (\phi_R-\phi_L)<0$, is such that (1)the fluids on the right expands un-affected into the left within its ``causal'' future; (2) $\phi$ and $\rho$ are constant along the rays emitting from the origin with their specific profile across the rays being determined by the free-streaming (4.33) 
of $e^{2 \phi}$ in the light-cone coordinates and  (4.28)
.
B. The right-moving rarefaction wave solution for $\rho_R-\rho_L=c_s (\phi_R-\phi_L)$. This corresponds to choosing the $+$ sign in (4.28)$-$(4.34)
.}
\end{figure}

Thus we have reached four one-parameter families of special solutions, two semi-infinite shock curves
\begin{equation}
\rho_R-\rho_L =-\frac{1}{2 \sigma} \text{log} \frac{[\sqrt{\sigma^2 \text{cosh}^2 (\phi_R-\phi_L)-(\sigma-1)^2}\mp\sigma \text{sinh} (\phi_R-\phi_L)]^2}{2 \sigma-1} \,\,\,\text{for}\,\,\, \phi_L>\phi_R, \label{sc}
\end{equation}
and two semi-infinite rarefaction-wave curves
\begin{equation}
\rho_R-\rho_L=\pm c_s (\phi_R-\phi_L)\,\,\,\,\,\,\,\text{for} \,\,\,\,\,\,\,\phi_L<\phi_R. \label{rc}
\end{equation}
The existence of the shock waves versus rarefaction waves for $\phi_L>\phi_R$ versus $\phi_L<\phi_R$ is in direct analogy to the case of Burgers (\ref{rwb1})-(\ref{matching}), though in the present case, they alone do not constitute the complete solution to the Riemann problem. On the other hand, the four semi-infinite curves emitted from the point $(\phi_L, \rho_L)$ divide the $(\phi,\rho)$ plane into four regions and furnish it with a complete coordination (Figure 5.A). This suggests to generate generic solutions by gluing the special ones. Consider for example the region between the two shock-curves $s_{\pm, L}$ in Figure 5 A. If $(\phi_R, \rho_R)$ is represented by the point $R1$ as in the figure, we may construct the future evolution of this Riemann problem by gluing together a left-moving shock solution between $(\phi_L, \rho_L)$ and $(\phi_{M1}, \rho_{M1})$ followed by a right-moving shock solution between $(\phi_{M1}, \rho_{M1})$ and $(\phi_{R1}, \rho_{R1})$. The point $M1$ must then sit on the left-moving shock curve from $L$ (represented by $s_{-,L}$ in Figure 5.A), and its right-moving shock curve $s_{+, M1}$ must pass through the point $R1$. This determines the point $M1$ uniquely as in Figure 5.A. The resulting solution is shown in Figure 5.B. Now consider an observer co-moving with the fluid in the middle region $M1$. As the relations $\rho_{M1}> \rho_L$ and $\rho_{M1}>\rho_{R1}$ show, she is on the downstream sides of both shocks and sees the two shocks moving apart from each other. This, combined with the homogeneity of the region $M1$ means that this gluing procedure produces a mathematically well-defined weak solution. Furthermore every fluid parcel's entropy increases every time it crosses a shock front, as is guaranteed by our choice (\ref{riemann1}). Hence the solution is physically sensible. For the other three regions in the $(\phi, \rho)$ plane, similar analysis applies. And the rule is summarized as follows: (1) In the special case that $(\phi_R, \rho_R)$ lies on one of the four semi-infinite curves $s_{\pm, L}$, $r_{\pm, L}$ from $(\phi_L, \rho_L)$, we have a simple future evolution consisting of a single shock or rarefaction wave. (2) In the generic case, $(\phi_R, \rho_R)$ falls into one of the four bulk regions with respect to $(\phi_L, \rho_L)$. Then the solution consists of two objects, shock wave-shock wave, shock wave-rarefaction wave, rarefaction wave-rarefaction wave, or rarefaction wave-shock wave, which combination occurs depends on the region $(\phi_R, \rho_R)$ falls in. The intermediate point $M$ is universally determined by the rule that, it must sit on the proper $-$ curve from $L$ ($r_{-,L}$ or $s_{-,L}$), and the proper $+$ curve from $M$ ($r_{-,M}$ or $s_{-,M}$) must pass through the point $R$ (Figure 5)\footnote{We note in pass that, for any point $p$ in the $(\phi, \rho)$ plane, (\ref{softshock}) and (\ref{s3}) together imply that the curves $s_{-,p}$ and $r_{-,p}$ are connected in at least a $C^1$ fashion at the point $p$, as are the curves $s_{+,p}$ and $r_{+,p}$. }.
\begin{figure}[h!]
\vspace{0 cm}
\begin{center}
\scalebox{2}[2]{\includegraphics[width=0.5\textwidth]{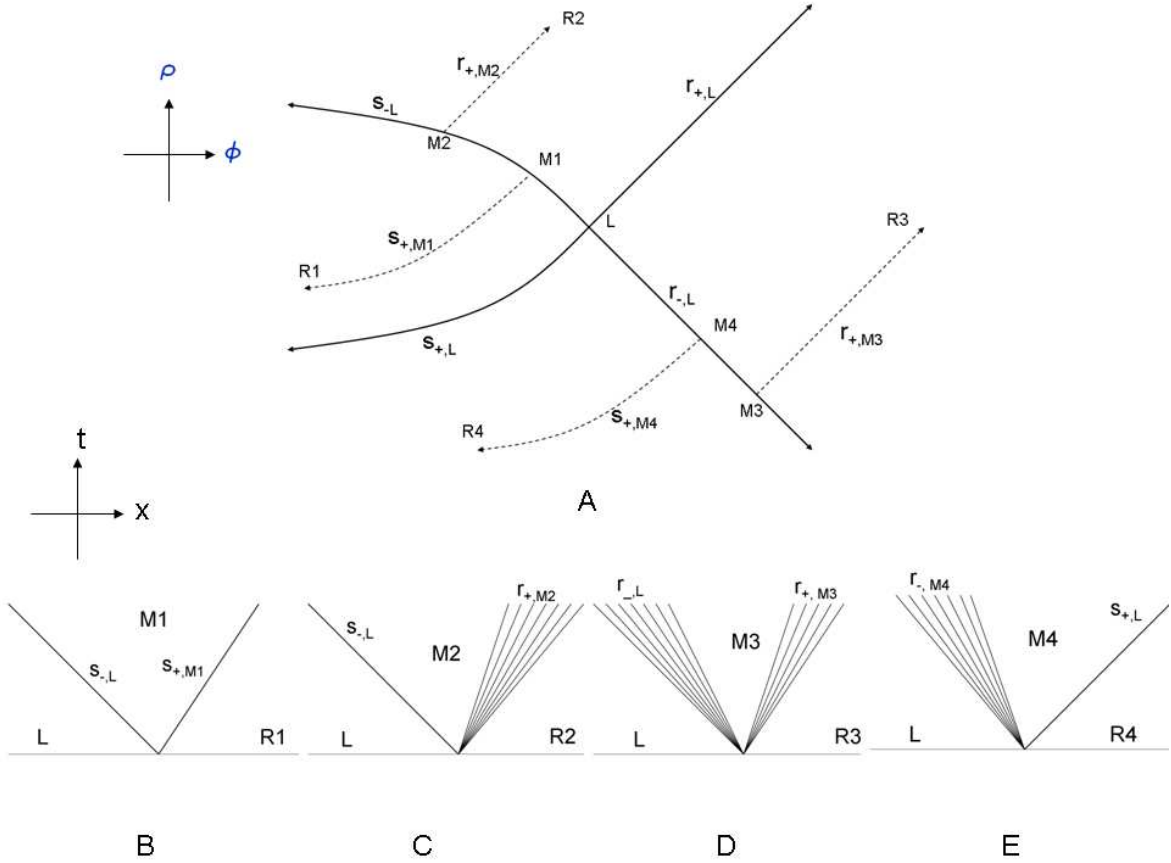}}
\end{center}
\vspace{0 cm}
\caption{A. The shock and rarefaction curves through various points in the $(\phi,\rho)$ plane. The notation is such that $r_{\pm ,p}$ represents the right/left-moving rarefaction curve through the point $p$ in the plane, and that, similarly, $s_{\pm ,p}$ represents the right/left-moving shock curve through $p$. The position of the middle point $M$ is determined by the fact that it needs to sit on the proper $-$ curve through $L$, and the proper $+$ curve through $M$ must pass through the point $R$. B. Solution for the Riemann problem $(L, R1)$: $s_{-}(L, M1) \bigsqcup s_{+}(M1, R1)$. C. Solution for $(L,R2)$: $s_{-}(L, M2) \bigsqcup r_{+}(M2, R2)$. D. Solution for $(L, R3)$: $r_{-}(L, M3) \bigsqcup r_{+}(M3, R3)$. E. Solution for $(L, R4)$: $r_{-}(L, M4) \bigsqcup s_{+}(M4, R4)$. }
\end{figure}

\subsubsection{Scattering between shock fronts}

Now equipped with the complete solution to the Riemann problem (\ref{left}) and (\ref{right}) of (\ref{rone}) and (\ref{rtwo}), we proceed to classify the shock-scattering processes. The basic idea is, in analogue to Burgers, to exploit the future uniqueness and past non-uniqueness of the physical solutions to the Riemann problem. The results for $2 \rightarrow 2 $ scattering are shown in Figure 6, which, together with the captions, are self-explanatory.

We first collect the physics of the case when the two colliding shocks are strong: $- \delta \phi_i=- (\phi_{R,i}-\phi_{L,i}) \gg 1$. If the incoming shocks travel in opposite directions (in the rest frame of the fluid between them) collide, they essentially pass through each other with only exponentially small modifications to their amplitudes on the order of $e^{- 2 |\delta \phi_i|}$. The major consequence of the collision is in stead a large increase of temperature and a boost of the fluid in the region in-between:
\begin{eqnarray}
\label{middle1}
\rho_{M'}-\rho_M &=& \frac{1}{\sigma} (| \delta \phi_1| + | \delta \phi_2|)+ \frac{2}{\sigma} \text{log} \frac{\sigma^2}{2 \sigma-1}+\circ (e^{- 2 |\delta \phi_i|}) \gg 1, \label{m1}\\
\phi_{M'}-\phi_M &=& |\delta \phi_1|-|\delta \phi_2|. \label{m2}
\end{eqnarray}
When two shocks traveling in the same direction (in the rest frame of the fluid between them) catch on each other, the collision glues them into a single shock whose strength is essentially the sum of the two incoming ones. The space between the original shocks disappears, and the single outgoing shock interpolates between the two outer regions before the collision, except up to a modest rarefaction wave emitted in the opposite direction (see Figure 6c for example). The strength of this rarefaction wave is of order 1 and is independent of the strength of the colliding shocks in the strong shock limit. This means that the new region between the outgoing rarefaction wave and the shock wave (M2' in Figure 6c) only mildly defers from the region L, and differs very strongly from the region M2, in terms of both a large increase in temperature and a large boost towards the direction of the motion of the shocks.

In the generic case of not necessarily strong, 2 or more-than-2 shocks scatterings, it is clear from the solution of the Riemann problem that no more than $2$ products come out, and the result is, as before, uniquely encoded by the physical conditions in the outermost regions $(\phi_L, \rho_L)$ and $(\phi_R,\rho_R)$. Complete information for the scattering is readily extracted from the equations for the curves (\ref{sc}) and (\ref{rc}).
\begin{figure}[h!]
\vspace{0.0 cm}
\begin{center}
\scalebox{2}[2]{\includegraphics[width=0.5\textwidth]{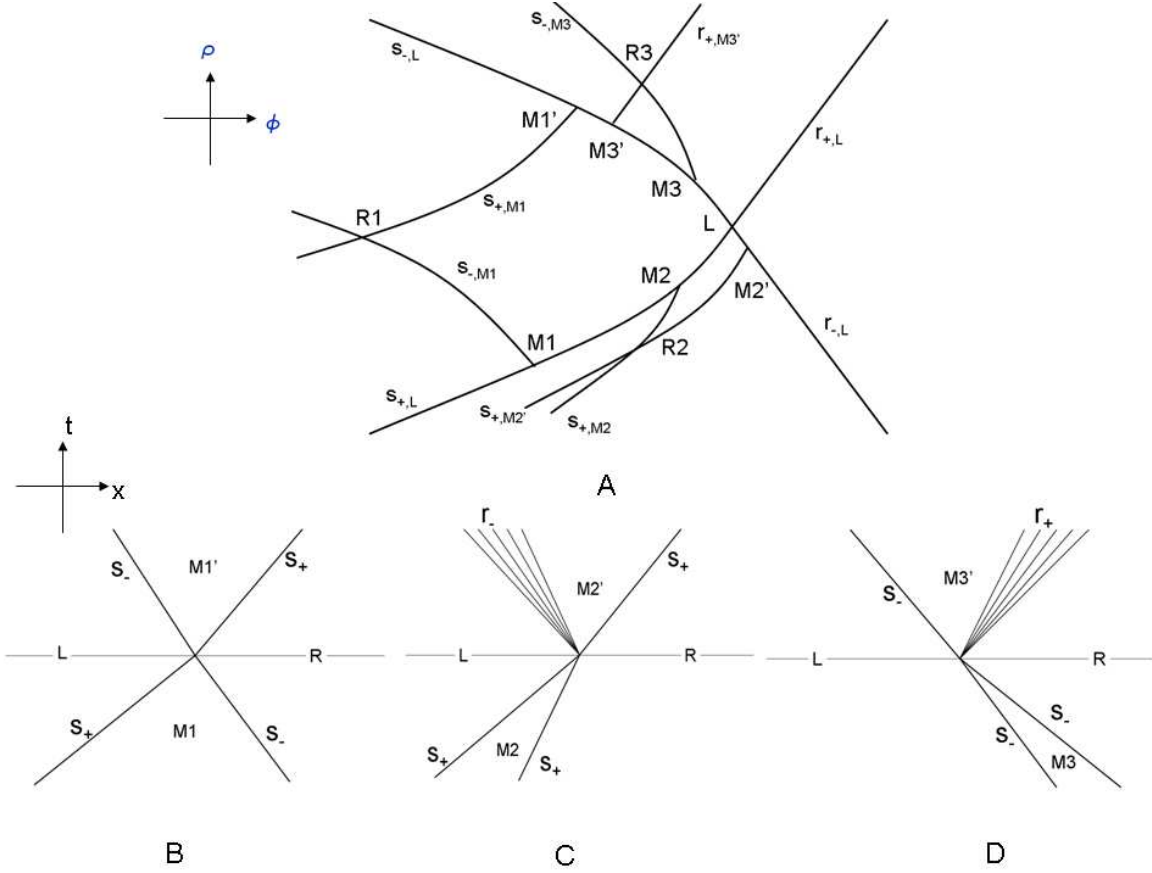}}
\end{center}
\vspace{0 cm}
\caption{A. The shock and rarefaction curves through various points in the $(\phi,\rho)$ plane. The notation is as in Figure 5. The task is to determine the positions of $M_i'$ in terms of $L$, $M_i$, and $R_i$. In the present $2 \rightarrow 2$ scattering, the position of $M_i$ is uniquely determined by that of $L$ and $R_i$. And therefore $L$ and $R_i$ uniquely determine $M_i'$. The results are shown in the next three figures. It is clear that in general scattering with more than two incoming objects, there still exist generically, and at most two outgoing objects. And their properties are again uniquely determined by the two outermost regions $L$ and $R$.  B. Head-on shock collision: $s_+ + s_-\rightarrow s_-+s_+$.  C. Catch-on shock collision: $s_++s_+ \rightarrow r_-+s_+$. D. Catch-on shock collision: $s_-+s_- \rightarrow s_-+r_+$. }
\end{figure}
\bigskip

\subsubsection{Scattering between shocks and simple waves}

We proceed to compute the outcome of the collision between a shock wave and a simple wave. This is the generalization of the shock-sound scattering of section 4.1 to the nonlinear regime. We mostly discuss the case of head-on collision of a simple wave on the shock front, where only transmission wave but not reflection wave is expected to arise. The key assumption we make is that the transmission wave is again a simple wave, which, as we presently see, is self-consistent. The spacetime physics is illustrated in Figure 7a for the collision between a right-moving simple wave and a left-moving shock wave; the other case of head-on collision of a left-moving simple wave off a right-moving shock wave is identical up to space reflection.

Note that the time-like interval to the immediate left of the shock trajectory during the course of the collision is mapped to the right-going rarefaction curve connecting $A$ and $B$ in the $(\phi, \rho)$ plane, the interval to the immediate right of the shock is mapped to the right-going rarefaction curve through $C$. Since the two curves are parallel in our case, we infer that the shock strength ($\Delta_s\equiv\Delta \phi_{\text{shock}}=\phi_{L,\, \text{shock}}-\phi_{R,\,\text{shock}}>0 $) remains constant during the collision, and that the overall amplitude of the simple wave (in terms of $\Delta \phi_{\text{simple wave}}=\phi_{R,\, \text{simple}}-\phi_{L,\, \text{simple}}>0$) also remains un-modified after the collision, the latter fact being the generalization of (\ref{transm1}) into the nonlinear regime. This also determines the property of the middle region $D$ after the collision, as shown in the figure
\begin{equation}
(\phi_D, \rho_D)-(\phi_A,\rho_A)=(\phi_C,\rho_C)-(\phi_B,\rho_B).
\end{equation}

The deflection of the simple wave past the shock front is determined by (\ref{s4}) (choosing $c_+$) and the value of $\phi$ along the time-like interval to the immediate right of the shock locus. The trajectory of the shock during the collision can be determined using (\ref{match4}), taking the $-$ sign for the left-going shock
\begin{eqnarray}
\label{shock trajectory}
e^{2 \phi_s} &=& e^{\phi_L+\phi_R} \frac{(1+c_s^2) \text{cosh} \Delta_s-\sqrt{(1+c_s^2)^2 (\text{cosh} \Delta_s)^2-(1-c_s^2)^2}}{1-c_s^2} \label{t1}\\
&=& e^{2 \phi_L-\Delta_s} \frac{(1+c_s^2) \text{cosh} \Delta_s-\sqrt{(1+c_s^2)^2 (\text{cosh} \Delta_s)^2-(1-c_s^2)^2}}{1-c_s^2}. \label{t2}
\end{eqnarray}
Here $\phi_L (x)$ takes its value on the rarefaction curve $AB$, and its $x$-dependence is specified by the particular simple wave solution participating in the interaction. Take the right-moving centered rarefaction wave as an example (Figure 7), it describes the free-streaming of $e^{2 \phi}$ in the light-cone (\ref{s5})
\begin{equation}
e^{2 \phi_L} = \frac{x^+}{c_+ x^-}=\frac{1}{c_+} \frac{t+x}{t-x}. \label{phiL}
\end{equation}
Feeding (\ref{t2}) and (\ref{phiL}) into $$\frac{\text{d} x_s}{\text{d} t}=\text{tanh} \phi_s$$ and integrate, we find
\begin{equation}
(t+x_s)^{a+1}(t-x_s)^{a-1}=\text{Const},
\end{equation}
where $0<a<1$ is
\begin{equation}
a(c_s, \Delta_s)= -\frac{[(1+c_s^2) \text{cosh} \Delta_s-\sqrt{(1+c_s^2)^2 (\text{cosh} \Delta_s)^2 -(1-c_s^2)^2}]-(1+c_s)^2 e^{\Delta_s}}{[(1+c_s^2) \text{cosh} \Delta_s-\sqrt{(1+c_s^2)^2 (\text{cosh} \Delta_s)^2 -(1-c_s^2)^2}]+(1+c_s)^2 e^{\Delta_s}}.
\end{equation}
Recall $\Delta_s \equiv \phi_L-\phi_R>0$ is the constant strength of the shock.

\medskip

We note the stability of both the shock and the rarefaction wave under this particular collision process, which follows directly from the translation invariance in the $(\phi, \rho)$ plane of the rarefaction and shock curves. Physically, the incoming simple wave is a large, negative temperature disturbance to the shock front from the upstream side. It enhances the temperature difference between the two sides of the shock, and hence it is not surprising that it does not weaken the shock front.

\medskip

The case of catch-on collision between a simple wave and a shock wave traveling in the same direction is more complicated to compute. Due to the appearance of reflection wave, simple wave ansatz is expected to break down in the region where incoming and reflection waves overlap. We merely note that, in this case, the simple wave impinges on the shock front as a large, negative temperature rarefying disturbance from the downstream, and hence at least partially, it cancels the discontinuities at the shock front. Thus we do expect a significant cancelation between the shock and the simple wave in this case.

\begin{figure}[h!]
\vspace{0.0 cm}
\begin{center}
\scalebox{2}[2]{\includegraphics[width=0.5\textwidth]{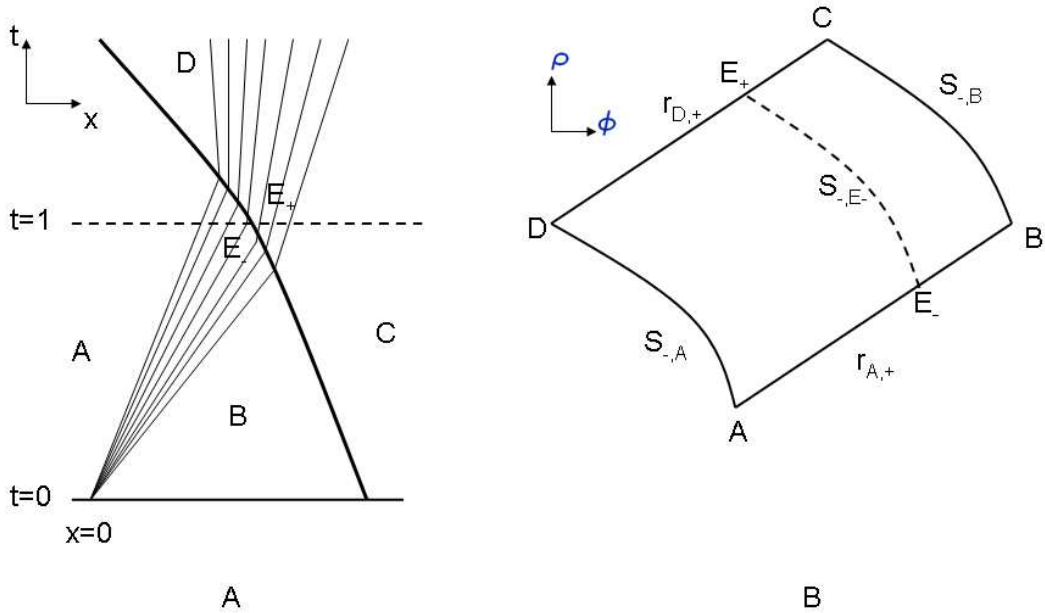}}
\end{center}
\vspace{-2.7 cm}
\caption{A. The spacetime diagram of a collision between a rarefaction wave $r_+$ and $s_-$. The shock penetrates through the rarefaction wave, with a slight, continuous change of direction. The rarefaction wave deflects its motion across the shock. Each wave maintains its strength, a consequence of the translation invariance in the $(\phi, \rho)$ plane of the rarefaction and shock curves. B. The collision process in the $(\phi,\rho)$ plane. Each constant time slice is mapped into an open contour in the $(\phi, \rho)$ plane from $A$ to $C$:  $t=0$, $A \rightarrow B \rightarrow C$; $t=1$, $A \rightarrow E_- \rightarrow E_+ \rightarrow C$; $t=+\infty$, $A \rightarrow D \rightarrow C$.}
\end{figure}

\subsection{An Effective Field Theory of Shocks and Sounds? }

We pause to summarize the lessons from the previous two subsections.
Based on a perturbative analysis of the junction condition across the shock front, as well as the complete solution to the Riemann problem, we worked out quantitatively almost a complete set of interaction vertexes in various scattering channels involving shock waves.

For shock-sound scattering, we worked out the transmission and reflection amplitudes of the sound wave, as well as its backreaction on the shock front. We specifically ruled out the perturbative instability of the shock front towards spontaneous emission of sound waves.

For shock-shock scattering, take the limit of strong shocks as an example. When the two shocks collide head-on, they pass through each other almost un-disturbed, leaving behind the region between them drastically heated up. As for a catch-up collision, the two shocks glue together to form a single, stronger shock with essentially additive strength (with small corrections) and continue the movement.

For the case of head-on collision between a shock wave and a simple wave, the shock front maintains its strength and continuously accelerates/deccelerates (depending on the relative direction of the waves and the frame of reference) until the end of the encounter.

In these above collision channels, the shocks either maintain their amplitudes or strengthen additively so that the total strengths of the colliding shocks $\sum_i |\delta \phi_{s,i}|$ and $\sum_i |\delta \rho_{s,i}|$ are approximately conserved after the collision. This (approximate) conservation is nontrivial in cases that rarefaction waves are involved in the incoming and/or outgoing states.

In the case that a simple wave impinges on the shock front with opposite properties to that of the shock front (i.e. with cancelation signs for both the temperature and rapidity discontinuities sourced by the shock front), such as the case when a simple wave incomes from the downstream, we do expect significant weakening of the shock, if not complete annihilation. However, it is technically more challenging to make computation in this case.

\medskip

It is very tempting to construct an effective interacting field theory consisting of shock waves, sound waves, and rarefaction waves, based on the scattering data we computed in this section. The shock waves and the rarefaction waves behave as solitons and anti-solitons with chiralities and approximate conservation properties, on top of which propagates the sound wave. The interaction vertexes should also be chirality-dependent\footnote{It appears interesting to understand, in this context, whether the dissipative anomaly of the hydrodynamics may be represented as the anomaly of a global symmetry of the effective field theory. The functional integral of the latter setup originates from the integration over the proper statistical ensemble of the external force and that of the initial conditions in the former setup.}. The extent to which such a model may effectively captures certain features of the relativistic hydrodynamic turbulence is not clear. However, given the complexity of the turbulence phenomenon, the very limited understanding that is available, and the theoretical difficulty to reach further, one may hope a model of this nature to shed much light on the complex dynamics of turbulence and the transition to turbulence from laminar flows. Hence we hope to return to this problem in the future.

\section{Singularities and Statistics, Blowups and Preshocks}


In section 3, we focused on shock fronts and their contributions to the short distance limit of two-point functions of turbulent flows. They are but a special case of a simple, yet general relationship that exists, in the in-viscid limit, between various singular loci of hydrodynamic flows and various limits of statistical properties of turbulence. While knowledge of such singular loci are far from enough for a complete characterization of statistical properties of turbulence, they do capture various limits of the statistics in a direct way, in that only in the vicinity of the proper singularities, the corresponding quantities may develop large, unbounded values.

As an example other than the two-point functions we considered thus far, consider the question of determining the behavior of the probability density function of the velocity gradient in the regime where it takes very large, negative values. In the case of unforced Burgulence, the answer
\begin{equation}
\label{pdf}
p(\xi)\propto |\xi|^{-\frac{7}{2}},\,\,\,\,\,\,\,\,\,\,\,\,\,\,\,\,\,\,\,\,\text{for}\,\,\, \xi\rightarrow- \infty
\end{equation}
where $\xi=\frac{\partial u}{\partial x}$, was derived by noting that, neither can large, negative velocity gradients come from smooth components of the flow and regions close to matured shock fronts, at both places $\xi$ is finite, nor can they come from the matured shock loci where $\xi$ is divergent; they only arise close to pre-shocks, locations at which shocks are initially formed. A close study of the Lagrangian map furthermore shows that, to have $\frac{\partial u}{\partial x}<\xi\ll0$, the spatial width of the region near the pre-shock scales as $|\xi|^{-\frac{3}{2}}$ and the temporal size scales as $|\xi|^{-1}$, and therefore the spacetime volume and hence cumulative probability to have $\partial_x u <\xi\ll0$ is proportional to $|\xi|^{-\frac{5}{2}}$, and hence the result (\ref{pdf}) for the probability density function.

\bigskip

A natural generalization of our work is therefore to classify all possible singularities in the relativistic hydrodynamical solutions and study the roles they play in determining the statistical properties of turbulent flows. While in the case of Burgers equation and its higher dimensional generalizations, the study of the dynamics of singularities of the solutions to the 1st order PDE's are accomplished via the study of the properties of a finite, smooth function in a certain multi-dimensional space, the Lagrangian potential $\varphi(a,t)$ in $1+1$d Burgers (\ref{lagangian potential}) or
\begin{equation}
\label{lag}
\varphi(\textbf{a},\,t)\equiv t \phi(\textbf{a},0) - \frac{|\textbf{a}|^2}{2}
\end{equation}
in higher d Burgers, the problem appears to be much more difficult in the case of the 2nd order PDE of the relativistic hydrodynamics. The first task is therefore to find the proper mathematical tools to analyze this question.  The problem of constructing and classifying (and possibly continuing beyond) singularities of solutions to 2nd order hyperbolic PDE's appears to be a difficult and lasting research direction in mathematics itself that we, during the course of this project, scratched upon.

\bigskip

In the absence of a systematic study of the possible singularities, we make two simple remarks. The first concerns the question, under time evolution from a reasonable initial state, can a hydrodynamic solution develop a local divergence (i.e. going to infinity) in the value of the temperature $\rho$ at a finite time? Were such divergence to develop, what are the rules for extending the solution beyond such singularities? For initial states very close to constancy, theorems guaranteeing the long time existence of smooth solutions have been proven for the 2-by-2 genuinely nonlinear hyperbolic systems of conservation laws to which our system, in the in-viscid limit, belongs. The proof is based on carefully bounding the magnitudes of the solution and its various derivatives {\cite{decay}}. Such theorems are not helpful for our purpose, because (1) the initial states relevant for us are not necessarily close to constancy, and (2)we would also like to include the effect of external forcing. So the questions still remain open given these theorems.

At the level of the ideal hydrodynamics that we are presently concerned about, a local divergence of temperature would occur at a point $P$ if an infinite number of shocks of non-infinitesimally small amplitudes collide in the finite spacetime region of the causal past of $P$ after the initial time. Such a possibility may seem unlikely, but is not easily ruled out. Even if there is only a finite number of shocks on the initial time slice in the causal past of $P$, and hence only a finite number of collisions among themselves in the finite region bounded by the initial time slice and the backward light rays through $P$, an un-known amount of shocks may be produced by the smoother components of the fluid and by the external forcing inside this finite region. Whether these can result in a local temperature blow up, a priori, depends on the small scale spectral property of the initial state as well as that of the driving force.

What does seem plausible is that a local divergence of this nature would not arise if we turn any positive viscosity, as diffusion would broaden the shock width and make any finite spacetime region able to accommodate only a finite number of collision events. This suggests that, at the level of ideal hydrodynamics, even were finite time blow-up of local temperature to arise, there would exist an operational way to extend the solution beyond the blowups to construct sensible solutions, as the ideal, in-viscid limit should be described as the instantaneous local equilibration limit of a viscous fluid \footnote{There is indication, based on holographic/perturbative computations of transport properties of strongly/weakly coupled quantum field theories, that the straightforward in-viscid hydrodynamic limit is not physically realizable in relativistic quantum field theories. }.  On the other hand, we must note that when non-vanishing viscosity and hence positive mean free length is turned on, the very strong shocks, once formed, can no longer be described precisely by the viscous hydrodynamics if their thickness are smaller than the mean free length of the fluid particles. Aspects of the underlying microscopic theory is then needed for the further development of the solution. It is therefore a logical possibility that the relativistic hydrodynamics in the in-viscid limit would inherit this ``UV dependence'' in one way or another, possibly through the operational way through which to extend its solution beyond the finite time blowup singularities.


\medskip

We are, to a limited extent, able to construct a subset of possible singularities of solutions to the relativistic hydrodynamic equations other than the shocks.
Recall that under the simple wave ansatz (\ref{simple}), our system simplifies into either one of two Burgers systems (\ref{s5}) under the conditions (\ref{s3}). This simplification enables us to borrow all the knowledge about the solutions to the Burgers equation to the relativistic context \footnote{Note that this simplification of the simple wave dynamics only holds at the ideal level. When non-vanishing viscosity is turned on, the dynamics of the simple wave is much more involved. Here we again make the assumption that the correct simple wave dynamics in the in-viscid limit of the relativistic hydrodynamics is directly captured by the analysis carried out at the level of ideal hydrodynamics, combined with a dissipative constraint to the effect of selecting the (maximal) entropy increasing type of weak solutions whenever non-uniqueness arises. }. This includes the description of arbitrary solutions of the simple wave type in terms of the Lagrangian potential $\varphi_{\pm}(x^+_0,\,x^-)$, and specifically, the description of formation of shocks in terms of a transition from a locally convex $\varphi_{\pm}(x^+_0,\,x^-)$ for $x^-<x^-_\ast$ to a locally concave $\varphi_{\pm}(x^+_0,\,x^-)$ for $x^->x^-_\ast$. Among other things, this implies, similar to (\ref{pdf}), that there exists a power-law tail for the probability density function of the gradient of the rapidity $\phi$ when it takes large, negative values. This, in turn, implies a power law tail in the probability density function of the gradient of the temperature $\rho$ near both the positive and the negative infinity, arising from the left and right moving types of the simple waves. However, given that we lack a complete understanding of all possibilities, we do not know if such power-law tails are buried by any other contributions not of the simple wave type.

\medskip

To the extent that the sharp features (shocks, pre-shocks, weak discontinuities, etc) in the hydrodynamic solutions are smoothed away by turning on finite viscosity, so is their distinct contribution to the relevant limits of the turbulence statistics.

\bigskip
\bigskip
\centerline{\bf{Acknowledgements}}
\medskip
X.L. would like to thank Steven Schochet for helpful conversations.
The work is supported in part by the Israeli
Science Foundation center of excellence, by the Deutsch-Israelische
Projektkooperation (DIP), by the US-Israel Binational Science
Foundation (BSF), and by the German-Israeli Foundation (GIF).


\newpage

\end{document}